\documentclass[article]{aa} 

\usepackage{graphicx,natbib,txfonts}
\usepackage{longtable}
\usepackage{xcolor}
\usepackage{amssymb}
\usepackage{hyperref}

\def\Tcmb{\hbox{$T_\mathrm{CMB}$}}

\def\nH2{\hbox{$n_\mathrm{H_2}$}}
\def\kms{\hbox{km\,s$^{-1}$}}
\def\PKS1830{\hbox{PKS\,1830$-$211}}
\def\cm-2{\hbox{cm$^{-2}$}}
\def\fH2{\hbox{$f_{\rm H2}$}}
\def\fc{\hbox{$f_c$}} 
\def\nH{\hbox{$n_{\rm H}$}}

\def\Rq{\hbox{$\mathcal{R}_q$}}
\def\RXX{\hbox{$\mathcal{R}_{\rm XX}$}}
\def\RYY{\hbox{$\mathcal{R}_{\rm YY}$}}
\def\Rpol{\hbox{$\mathcal{R}_{\rm pol}$}}
\def\absRpol{\hbox{$\left | \mathcal{R}_{\rm pol} \right |$}}

\begin{document}

\title{Cosmo-tomography toward \PKS1830: Variability of the quasar and of its foreground molecular absorption monitored with ALMA}

\author{S.~Muller \inst{1}
\and I.~Mart\'i-Vidal \inst{2,3} 
\and F.~Combes \inst{4} 
\and M.~G\'erin \inst{5} 
\and A.~Beelen \inst{6} 
\and C. Horellou \inst{1} 
\and M.~Gu\'elin \inst{7} 
\and S.~Aalto \inst{1}  
\and J.\,H.~Black \inst{1} 
\and E. van Kampen \inst{8} 
}

\institute{Department of Space, Earth and Environment, Chalmers University of Technology, Onsala Space Observatory, SE-43992 Onsala, Sweden
\and Departament d’Astronomia i Astrof\'isica, Universitat de Val\`encia, C. Dr. Moliner 50, E-46100 Burjassot, Val\`encia, Spain
\and Observatori Astron\`omic, Universitat de Val\`encia, C. Catedr\`atico Jos\'e Beltr\'an 2, E-46980 Paterna, Val\`encia, Spain
\and Observatoire de Paris, LERMA, Coll\`ege de France, CNRS, PSL University, Sorbonne Universit\'e, Paris, France 
\and LERMA, Observatoire de Paris, PSL University, CNRS, Sorbonne Universit\'e, 75014 Paris, France
\and Aix-Marseille Universit\'e, CNRS \& CNES, Laboratoire d’Astrophysique de Marseille, 38, Rue Fr\'ed\'eric Joliot-Curie, 13388 Marseille, France
\and Institut de Radioastronomie Millim\'etrique, 300, rue de la piscine, 38406 St Martin d'H\`eres, France
\and European Southern Observatory, Karl-Schwarzschild-Str. 2, 85748 Garching b. M\"unchen, Germany
}

\date {Received  / Accepted}

\titlerunning{An ALMA monitoring of \PKS1830}
\authorrunning{Muller et al. 2023}

\abstract{Time variability of astronomical sources provides crude information on their typical size and on the implied physical mechanisms. \PKS1830\ is a remarkable radio-bright lensed quasar with a foreground molecular absorber in the lens galaxy at $z=0.89$. Small-scale morphological changes in the core--jet structure of the quasar ---which is magnified by the lensing--- result in a varying illumination of the absorber screen, which in turn causes variations in the absorption profile.}
{We aim to study the time variations of the system (the two main lensed images of the quasar and the two corresponding sightlines in the absorber) in order to obtain constraints on both the quasar activity and small-scale structures in the interstellar medium of the absorber.}
{We used ALMA to monitor the submillimeter continuum emission of \PKS1830, together with the absorption spectra of the H$_2$O and CH molecules, with 17 visits spread over six months in 2016. Complementing this, we used available ALMA data to investigate changes in the system in the period 2012-2022.}
{From the continuum data, we followed the evolution of the flux density, flux-density ratio, spectral index, and differential polarization between the two lensed images of the quasar; all quantities show significant variations related to the intrinsic activity of the quasar. We propose a simple parametric model of a core plus a ballistic plasmon to account for the continuum evolution, from which we constrain a time delay of $25 \pm 3$~days between main lensed images. 
The spectral lines reveal significant variations in the foreground absorption profile. A principal component analysis highlights apparent wavy time variations, possibly linked to the helical jet precession period of the  quasar.
From the deep averaged spectra towards the southwest image, we detect the absorption of the rare isotopolog $^{13}$CH and estimate an abundance ratio of $^{12}$CH/$^{13}$CH $\sim 150$. We also measure the oxygen isotopic ratios, $^{16}$O/$^{18}$O $= 65.3 \pm 0.7$ and $^{18}$O/$^{17}$O $= 11.5 \pm 0.5$ in the $z=0.89$ absorber. Finally, we find a remarkable continuous shallow trough in the water absorption spanning a velocity interval of nearly 500~\kms. This broad absorption could be the signature of an extra-planar molecular component.}
{All together, the system formed by the quasar \PKS1830\ and its foreground lens--absorber acts as a powerful gravitational microscope, providing us with the possibility to dissect small-scale structures in both the  ISM of the foreground absorber and the  jet of the background quasar.}


\maketitle

\section{Introduction}

Astronomical objects often appear to be of small angular size. Therefore, the study and characterization of these objects and of the inherent physical mechanisms (i.e., the whole subject of astrophysics) rely on our ability to access information corresponding to these small scales. This can be done in two ways: either with direct high-angular-resolution observations, for example by means of very long-baseline interferometry (VLBI), or, for varying sources, by studying their time variations and their timescales. Using the causality argument, the typical timescale of signal variations should indeed reflect the size of the corresponding regions.

The time variability technique has been applied to various sources and contexts, including compact transient sources like transiting exoplanets, supernovae, pulsars, and black holes, to name but a few. For example, the reverberation mapping technique within the broad-line region in active galactic nuclei (AGN) provides size estimates from the line variability, which, combined with spectroscopy measurements of the kinematics, yields a mass estimate for the central black hole (e.g., \citealt{bah72, bla82}). Another example is provided by the remarkable periodicity in the optical light curve of the gravitationally lensed quasar Q~J\,0158$-$4325, which with the help of microlensing led \cite{mil22} to propose a model based on a pair of supermassive black holes in the coalescence stage. Yet another application is the use of temporal variability of H\,I, optical, or ultraviolet absorption lines to study the small-scale structures down to a few AU in the interstellar medium (ISM; see e.g., a review by \citealt{sta18}).

The direction toward the lensed quasar \PKS1830, the subject of this work, is particularly favorable for studying time variability. \PKS1830\ is a blazar at $z=2.5$ \citep{lid99} with strong gravitational lensing due to an intervening nearly face-on spiral galaxy at $z=0.89$ \citep{wik96,win02}. The blazar shows significant activity on short timescales (down to hours and days) when observed from radio waves to X- and $\gamma$-rays (e.g., \citealt{abd15,mar19,mar20}). Two factors boost the apparent variability of \PKS1830  from our viewpoint on  Earth: {\em (i)} the orientation of the blazar's relativistic jet close to the line of sight, which makes it prone to apparent superluminal motions, and {\em (ii)} the lensing and its associated source magnification. As a result of the lensing geometry, the quasar appears as two main bright and compact images (to the northeast and the southwest, hereafter the NE and SW images), separated by $\sim 1\arcsec$ and embedded in an Einstein ring with a steep spectral index (mostly seen at centimetre (cm) wavelengths; e.g., \citealt{jau91}), and a weak third lensed image at approximately the position of the lens galaxy \citep{mul20b}. There is a geometrical time delay of $\sim 27$~days between the NE (leading) and SW images (see Sect.\,\ref{sec:timedelay}).

In addition to the peculiarities of \PKS1830\ mentioned above, the sightlines to the two main images intercept a molecular component in the disk of the $z=0.89$ lens galaxy, causing a remarkable molecular absorption-line system (\citealt{wik98, mul14}). A large number ($>60$) of molecules have been detected along the SW line of sight (e.g., \citealt{mul11,mul14,ter20}), which have been used not only to investigate the chemical and isotopic composition of the gas in the absorber but also as sensitive cosmological probes; for example, of the cosmic microwave background temperature \citep{mul13} and invariance of fundamental constants (e.g., \citealt{bag13,kan15,mul21}). However, these studies can be affected by systematic uncertainties due to the peculiarity of \PKS1830, with time variations being one of the major effects (e.g., \citealt{mul21}).

VLBI observations have revealed micro-changes in the morphology of the quasar on the timescale of months \citep{gar97}. Furthermore, the apparent separation between the two lensed images has been measured to vary by up to 200~$\mu$as
within eight months \citep{jin03}, which was interpreted by \cite{nai05} as evidence for recurrent plasmon ejections along a helical jet, with a jet precession period of one year. In addition, there is evidence for micro-lensing events in the X-ray and $\gamma$-ray light curves (e.g., \citealt{osh01,ner15}), implying that the hard emission is produced in the vicinity of the supermassive black hole.

These changes in the morphology of the quasar lead to variation in the illumination of the foreground material in the lens--absorber, and significant variation in the absorption profiles of molecular lines has been reported between spectra taken months to years apart at millimeter (mm) wavelengths \citep{mul08,sch15,mul14,mul21}. In contrast, there seems to be no line variability at cm wavelengths: \cite{all17} and \cite{com21} reported no significant variation in the redshifted H\,I 21-cm and OH lines between observations made at up to 20~year time intervals. This could be explained if the H\,I gas is distributed on larger scales than that probed by the changes in the source structure at these wavelengths. Regarding variations of absorption lines, the situation is indeed more favorable for molecular lines at high frequencies (i.e., in the mm/submm part of the electromagnetic spectrum): the background continuum emission is smaller (e.g., \citealt{gui99}) and originates closer to the base of the jet; in addition, the molecular screen is expected to be more clumpy and to have a smaller scale height than the atomic screen.

Overall, \PKS1830,\, with its sustained blazar activity, its low-inclination jet with superluminal motions, its lens magnification, potential micro-lensing events, and its intervening lens--absorber with absorption profile variability, is thus a particularly interesting source for time monitoring. In this work, we present the results of a monitoring of the continuum activity of \PKS1830  and of H$_2$O and CH submm absorption lines with the Atacama Large Millimeter Array (ALMA) over a time span of six months in 2016. The monitoring is complemented by ALMA archival data in the 2012-2022 period. Observations are presented in Sect.\,\ref{sec:obs}. Section\,\ref{sec:contvar} covers the quasar continuum variations. Results from spectral line absorptions are given in Sect.\,\ref{sec:linevar}. A discussion in Sect.\,\ref{sec:discussion} brings together the outcomes of continuum and spectral-line variations. A summary is given in Sect.\,\ref{sec:summary}.

\section{Observations} \label{sec:obs}

\subsection{2016 monitoring campaign}

Observations were made during ALMA Cycle 3, between March and September 2016. The proposed monitoring strategy comprised one visit every 10 days, over a time span of six months. To accommodate dynamic scheduling constraints, a time window of $\pm 7$~days was set for each visit. A journal of the observations is given in Table~\ref{tab:contdata-c3-ave}. The proposed cadence was achieved ---despite the scheduling challenge---, with a total of 17 successful visits. Among them, two visits were carried out on the same day (on 26 March 2016), providing a high signal-to-noise-ratio reference spectrum when combined, and the last visit (on  September 8 2016) was carried out after a gap of 42 days with respect to the penultimate visit. Otherwise, the time intervals between visits range between 6 and 21 days.

During the monitoring, the array configuration changed, but the longest baseline was always longer than 450~m, such that the synthesized beam was always smaller than 0.5$\arcsec$ and the two main lensed images of the quasar are therefore always well resolved. The on-source integration time for each visit was slightly less than 10~min, set by our original goal, which is to be able to track changes in the spectral line profiles down to a few percent during each individual visit.

Observations were made in dual polarization mode, with two orthogonal polarizers (XX and YY) in the antenna alt-azimuthal frame. The correlator setup was chosen to simultaneously cover the ground-state transitions of H$_2$O (557~GHz, rest frequency) and CH (two $\Lambda$-doublets at $\sim 533$ and 537~GHz, rest frequency) redshifted into Band~7. To cover these lines, three spectral windows were centered at $\sim$282.4, 284.6, and 295.4~GHz sky frequency, with a bandwidth of 938~MHz and a spectral resolution of 0.56~MHz, resulting in a velocity resolution of 0.6~\kms\ after Hanning smoothing. A fourth spectral window, not including any detectable spectral line, was centered at 296.8~GHz with a coarser spectral resolution of 31.25~MHz in order to provide extra continuum data.

The calibration was done with the CASA (version 4.7.2) package \citep{CASA2022}. The bandpass calibration was determined from the bright quasar J\,1924$-$2914. The flux calibration was carried out using Titan (Butler-JPL-Horizons 2012 model), Pallas, J\,1924$-$2914, or J\,1733$-$130, as indicated in Table~\ref{tab:contdata-c3-ave}. We expect an absolute flux accuracy of the order of 10\%-20\% according to the ALMA Technical Handbook \citep{cor22}. The gain calibrator was J\,1832$-$2039 in all visits. After the standard calibration, a step of phase-only self-calibration on \PKS1830\ was done with a short solution interval of 10~s.

The calibrated visibilities were finally fitted using the Python task UVMultiFit \citep{mar14} with a source model consisting of two point-like components, where the flux density of the NE image and the flux-density ratio, $\Re = f_{\rm NE}/f_{\rm SW}$, were set as free parameters. Different fit runs were carried out using Stokes~I data and data from the XX and YY polarizers separately in order to extract differential polarimetric information (see Sect.\,\ref{sec:Rpol}). The additional weak features of \PKS1830\ reported from ALMA observations by \cite{mul20b}, namely the third image (about 150 times weaker than the NE and SW images) and the small extensions from the NE and SW images (corresponding to the start of the Einstein ring and with a steep spectral index), have a negligible impact on the fitting process and were therefore not considered.


\begin{table*}[ht]
\caption{Journal of the 2016 ALMA monitoring and continuum data.} \label{tab:contdata-c3-ave}
\begin{center} \begin{tabular}{cccccccccrcc} \hline
Date & MJD $^{(a)}$  & N$_{\rm ant}$  & PWV $^{(c)}$ & Flux & PA $^{(d)}$ & $f_{\rm NE}$ $^{(e)}$ & $f_{\rm SW}$ $^{(e)}$  & $\mathcal{R}$ $^{(f)}$ & \Rpol $^{(g)}$ & $\Delta\alpha$ & $\sigma$ $^{(i)}$ \\
   & (-50000) & $^{(b)}$ & (mm) & calibrator & (deg)  & (Jy)  & (Jy)   &       &   ($\times 10^{-3}$)    &$^{(h)}$& (\%)        \\
\hline
2016/03/05 & 7452.545 & 36 & 1.2 &         Titan & 101 & 1.17 & 1.06&1.100\,(3) & $ 44.3\pm  1.3$ & 0.146(7)  & 0.49\\
2016/03/26 & 7473.385 & 42 & 1.0 & J\,1924$-$292 & 261 & 0.88 & 0.85&1.029\,(1) & $-12.5\pm  0.6$ & 0.027(7)  & 0.49\\
2016/03/26 & 7473.411 & 42 & 1.0 & J\,1924$-$292 & 259 & 0.91 & 0.89&1.025\,(1) & $-14.7\pm  0.7$ & 0.032(7)  & 0.47\\
2016/04/03 & 7481.398 & 43 & 1.9 & J\,1924$-$292 & 258 & 0.82 & 0.78&1.052\,(1) & $-29.8\pm  0.1$ & 0.042(9)  & 0.59\\
2016/04/10 & 7488.438 & 40 & 0.8 &         Titan & 103 & 0.80 & 0.78&1.027\,(1) & $ 15.5\pm  1.1$ & 0.039(8)  & 0.49\\
2016/04/16 & 7494.526 & 37 & 0.9 &        Pallas & 102 & 0.84 & 0.84&1.004\,(1) & $ 21.9\pm  0.8$ & 0.017(10) & 0.73\\
2016/04/23 & 7501.262 & 40 & 0.8 &         Titan & 260 & 0.71 & 0.73&0.967\,(1) & $  3.9\pm  0.4$ & 0.046(9)  & 0.57\\
2016/05/02 & 7510.258 & 39 & 0.8 &         Titan & 260 & 0.71 & 0.69&1.021\,(1) & $ 20.6\pm  0.4$ & 0.073(10) & 0.64\\
2016/05/08 & 7516.430 & 41 & 1.5 &        Pallas & 100 & 0.68 & 0.61&1.109\,(1) & $ 50.5\pm  0.4$ & 0.072(11) & 0.77\\
2016/05/17 & 7525.258 & 41 & 1.3 & J\,1924$-$292 & 260 & 0.64 & 0.55&1.146\,(1) & $ 44.4\pm  0.4$ & 0.065(10) & 0.64\\
2016/06/02 & 7541.442 & 37 & 1.3 &        Pallas & 105 & 0.57 & 0.54&1.053\,(1) & $ -0.4\pm  1.5$ & 0.069(28) & 2.14\\
2016/06/15 & 7554.163 & 38 & 0.6 & J\,1924$-$292 & 261 & 0.64 & 0.67&0.963\,(1) & $-35.5\pm  0.4$ & $-$0.007(10)& 0.67\\
2016/06/22 & 7561.158 & 37 & 1.2 & J\,1924$-$292 & 260 & 0.61 & 0.58&1.044\,(1) & $-30.5\pm  0.5$ & 0.000(12) & 0.95\\
2016/07/01 & 7570.236 & 42 & 1.1 & J\,1733$-$130 & 100 & 0.56 & 0.50&1.110\,(1) & $-10.9\pm  0.5$ & 0.061(12) & 0.83\\
2016/07/14 & 7583.318 & 39 & 0.7 &        Pallas & 104 & 0.56 & 0.50&1.115\,(1) & $  1.7\pm  0.2$ & 0.059(15) & 0.99\\
2016/07/28 & 7597.153 & 45 & 0.8 & J\,1733$-$130 & 100 & 0.43 & 0.45&0.962\,(1) & $  0.6\pm  4.1$ & $-$0.031(9)& 0.68\\
2016/09/08 & 7639.162 & 39 & 0.6 & J\,1924$-$292 & 104 & 0.19 & 0.19&1.010\,(2) & $  2.9\pm  0.6$ & 0.111(21)& 1.16\\
\hline \end{tabular} \tablefoot{$a)$ Modified Julian day; $b)$ Number of 12m antennas in the array; $c)$ Average precipitable water vapor in the atmosphere during the observations; $d)$ Parallactic angle; $e)$ Flux densities of the NE and SW images, respectively. We adopt a conservative absolute flux accuracy of $\sim 20$\%; $f)$ Flux-density ratio between the NE and SW images, the number in parenthesis gives the uncertainty on the last digit; $g)$ Polarization ratio, as defined in Eq.\,\ref{eq:rpol}; $h)$ Spectral index difference between the NE and SW images: $\Delta\alpha=\alpha_{\rm NE}-\alpha_{\rm SW}$; $i)$ rms noise level of the spectrum between 295.0 and 295.4~GHz (i.e., a line-free region in the spectral window centered on the redshifted H$_2$O transition), in \% of the continuum level of the SW image, at the native spectral resolution of $\sim 0.6$~\kms.} \end{center} \end{table*}

\subsection{Complementary data}

In addition to the 2016 monitoring data, we also used observations of \PKS1830\ available in the ALMA archive to extract continuum information (flux density and flux density ratio) between 2012 and 2022, as well as previous ALMA observations of the same fundamental transitions of CH and H$_2$O as mentioned above, as well as of the rare isotopologs H$_2^{18}$O and H$_2^{17}$O (see Table~\ref{tab:ComplementarySpec}). The data reduction was done with the same method as described above. All the spectral line data have a velocity resolution of $\sim 1.1$~\kms\ or better.

\begin{table*}[ht]
\caption{Complementary ALMA observations of H$_2$O, CH, and of the rare water isotopologs, H$_2^{18}$O and H$_2^{17}$O.}
\label{tab:ComplementarySpec}
\begin{center} \begin{tabular}{cccccc}
\hline
Species & Date & MJD$^{(a)}$ & $\delta v$$^{(b)}$ & $\sigma$$^{(c)}$ & $f_{\rm SW}$$^{(d)}$\\
        &      & (-50000) & (\kms)     & (\%)  & (Jy)    \\
\hline    

H$_2$O, CH & 2012/04/11   & 6028.8 & 1.0 & 1.1 & 0.6 \\
           & 2014/05/05   & 6782.8 & 0.5 & 1.3 & 0.3 \\
           & 2014/07/19   & 6857.8 & 0.6 & 0.5 & 0.6 \\
H$_2^{18}$O, H$_2^{17}$O  & 2014/05/05 & 6782.9 & 1.0 & 0.7 & 0.3 \\
H$_2^{18}$O, H$_2^{17}$O            & 2014/07/18 & 6856.7 & 1.1 & 0.3 & 0.6 \\
H$_2^{18}$O            & 2019/07/28 & 8692.1 & 1.1 & 0.1 & 1.7 \\
H$_2^{18}$O, H$_2^{17}$O            & 2022/08/19 & 9810.0 & 1.1  & 0.2 & 0.8 \\
\hline
\end{tabular}
\tablefoot{ $a)$ Modified Julian day; $b)$ Velocity resolution after Hanning smoothing; $c)$ rms sensitivity as a percentage of the SW image continuum level; $d)$ Flux density of the SW image.}
\end{center} \end{table*}

\section{Results from continuum data} \label{sec:contvar}

\begin{figure}[h!] \begin{center}
\includegraphics[width=8.2cm]{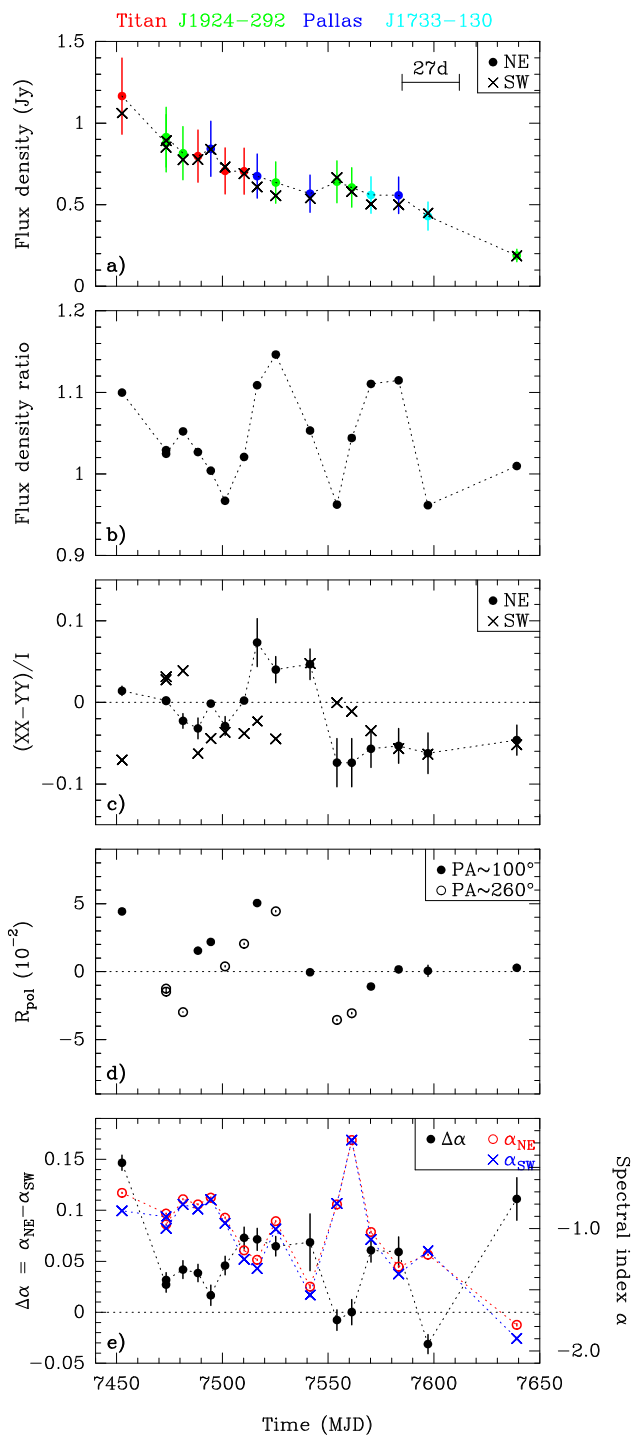} 
\caption{Time evolution of the continuum properties of \PKS1830\ during our monitoring in 2016 (see Table~\ref{tab:contdata-c3-ave}): $(a$) Flux density of the NE and SW images (error bars of 20\% are indicated for the NE image data points only; the color code indicates the source used for flux density calibration). $(b$) NE/SW flux density ratio. ($c$) Difference in flux density between the XX and YY polarizers normalized by the Stokes~I flux density. $(d$) Polarization ratio \Rpol\ (values are encoded as a function of the PA of the observations). $(e)$ Difference in spectral index. The expected time delay of $\sim 27$~days between the two lensed images of the quasar is indicated in panel $(a)$.}
\label{fig:cont-monitoring}
\end{center} \end{figure}

Table~\ref{tab:contdata-c3-ave} lists the values of the flux densities, flux density ratios, polarization ratios, and spectral index differences of the two NE and SW lensed images during the 2016 monitoring and we present their time variations in Fig.\,\ref{fig:cont-monitoring}. We used four different flux-density calibrators (Titan, Pallas, J\,1924$-$292, and J\,1733$-$130). The data points follow a relatively smooth light curve, which is free of significant outliers, giving confidence that potential systematic effects in the flux calibration are limited. In addition, while the absolute flux accuracy is of $\sim 10\%-20$\%, we emphasize that the accuracy of relative quantities, such as the flux-density ratio of both lensed images and the polarization ratio, is much higher (e.g., at a per mil level for the flux density ratio), and can therefore offer subtle diagnostics of the quasar activity, as discussed in the following subsections.

\subsection{Submillimeter light curve} \label{sec:LightCurve}

Over our six-month monitoring of \PKS1830, we observe a remarkable, nearly monotonic, and relatively featureless decrease in the flux density of the NE image, from $\sim 1.2$~Jy to $\sim 0.2$~Jy, which is a drop of more than 80\% of its flux density at the start of the monitoring. The evolution of the NE and SW images is highly correlated and on a longer timescale than the time delay between the two images, indicating that this spectacular drop is intrinsic to the quasar and not due to a differential lensing effect. On average, the flux density decreased by $\sim 5$~mJy/day over the six-month monitoring. This rate is small compared to  previously reported flux variations of up to $\sim 10$\% on an hourly timescale (e.g., \citealt{mar16,mar19}), but it is remarkable that it was sustained over such a relatively long period.

\subsection{Flux-density ratio} \label{sec:FluxRatio}

\begin{figure}[h] \begin{center}
\includegraphics[width=8.8cm]{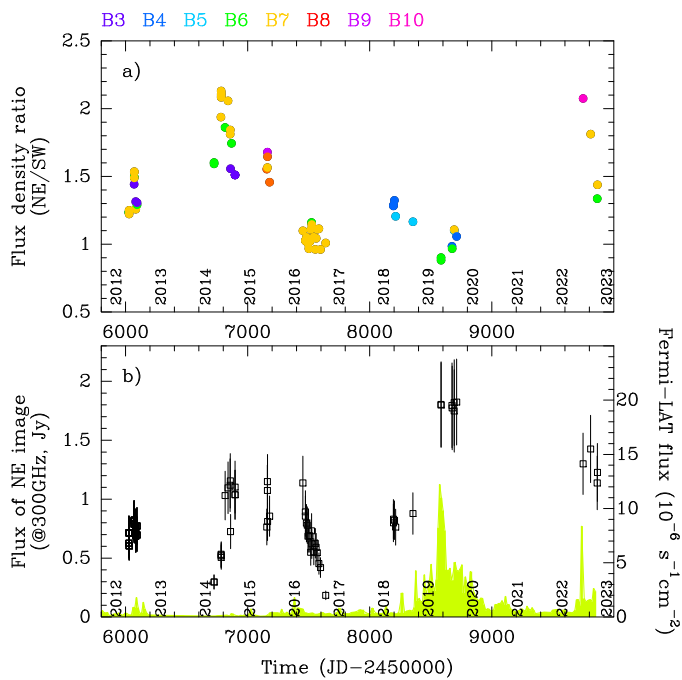}
\caption{History of ALMA measurements toward \PKS1830. {\em (a)} Flux-density ratios of \PKS1830\ NE and SW images (with the color code corresponding to ALMA bands, as indicated at the top of the figure).
{\em (b)} Flux densities of the NE image of \PKS1830, scaled to an equivalent frequency of 300\,GHz such as $S_{\rm 300GHz} = S_{\nu} \times (\nu/{\rm 300GHz})^{\alpha}$ with a spectral index $\alpha=-0.7$ (error bars indicate a nominal flux uncertainty of 20\%) overlaid on top of the Fermi-LAT $\gamma$-ray daily-monitoring light curve (in green).}
\label{fig:fluxratio-all}
\end{center} \end{figure}

In contrast to the large decrease in flux density during the 2016 monitoring, the instantaneous flux-density ratios $\Re = f_{\rm NE}/f_{\rm SW}$ show only mild ($\sim 10$\%) variations around an average value of $\sim 1.04$, with extreme values of 0.96 and 1.15 (Fig.\,\ref{fig:cont-monitoring}~b). The flux-density ratio measurements, being relative by nature, are free from instrumental effects. Given the high signal-to-noise ratio of the data, their accuracy is of the order of one per mil. As discussed by \cite{mar13}, the variations of the flux-density ratio over timescales of shorter than a few days to a few weeks are likely a direct signature of the intrinsic activity of the quasar modulated by the time delay between the two images, because we would expect timescales of longer than a month for milli-lensing.

The flux-density ratio is best measured at mm/submm wavelengths, where the contribution from the Einstein ring is lower due to its steep spectral index (\citealt{jau91}, see also \citealt{mul20b}). However, prior to ALMA it was difficult to obtain mm observations with sufficient angular resolution to separate the two lensed images (partly due to the southern declination of \PKS1830\ and to the small number of antennas of the previous generation of mm interferometers). Nevertheless, the flux-density ratio could be estimated from the saturation level of saturated lines, such as HCO$^+$ or HCN J=2-1, which block the SW image almost completely (e.g., \citealt{wik01,mul08}). Between 1995 and 2007, the flux-density ratio was found to vary between 1 and 3, but with only rare excursions to the extreme values. The average value of these old measurements was $1.7 \pm 0.3$ \citep{mul08}.

The flux-density ratios during the 2016 monitoring are among the lowest of all ALMA measurements since 2012, as shown in Fig.\,\ref{fig:fluxratio-all}a. In 2012, the first ALMA Cycle~0 observations were taken serendipitously at the time of a $\gamma$-ray flare. The submm flux-density ratios showed remarkable temporal and chromatic variations between 1.2 and 1.5 within two months, which were interpreted by \cite{mar13} with a simple model of plasmon injection and opacity effects in the jet of the  quasar. Subsequent ALMA observations in 2014--2015 show slightly higher flux-density ratios between 1.4 and 2.2, contrasting with the measurements in 2016--2020, which are closer to $\sim 1$, in particular those taken close in time to the record-breaking $\gamma$- and radio flare in 2019. The most recent measurements at hand (2022) show values in the range 1.3--2.0.

With the current collection of flux-density ratio measurements, it is difficult to obtain a clear and definitive determination of the true differential magnification, \Rq, between the two lensed images. Taken at face value, the average value of all the ALMA flux-density ratios measured between 2012 and 2022 is $1.38 \pm 0.35$ (with 1$\sigma$ dispersion) and the median is 1.29. However, the evolution of the flux-density ratios in Fig.\,\ref{fig:fluxratio-all}a shows a modulation on timescales of the order of several years, which could be the signature of ``milli-lensing'' (e.g., as discussed by \citealt{mar13}). On shorter timescales of a few days to weeks, micro-lensing could also affect the picture (e.g., \citealt{wam92}). This could be tested with a long and regular time monitoring, including multifrequency observations in order to probe the chromaticity of the flux-density ratio evolution.

\subsection{The time delay and the differential magnification factor} \label{sec:timedelay}

\begin{table*}[ht]
\caption{Previous measurements of the time delay $\Delta t$ and the differential magnification \Rq\ between the NE and SW lensed images.}
\label{tab:timedelay}
\begin{center} \begin{tabular}{ccccc}
\hline
Observations date & Method & $\Delta t$ (days) & \Rq & Note \\
\hline    
1990-1991 & Light-curve of VLA data at 8.4 and 15~GHz & $44 \pm 9$ $^{(a)}$ & $1.17-1.23$ & \cite{vanomm95} \\
1997-1998 & Light-curve of ATCA 8.6~GHz data & $26^{+5}_{-4}$ & $1.52 \pm 0.05$ & \cite{lov98} \\
1996-1998 & Saturation of the HCO$^+$ J=2-1 line & $24^{+5}_{-4}$ & -- & \cite{wik01} \\
1997 Jan-Apr & 43~GHz VLBI observations $^{(b)}$ & -- & $1.13 \pm 0.61$ & \cite{jin03} \\ 
2008-2010 & $\gamma$-ray light curve from Fermi-LAT & $27 \pm 0.6$ $^{(c)} $& -- & \cite{bar11}  \\
2012 Apr-Jun & Model of a chromatic jet to ALMA data & 27 $^{(d)}$ & 1.35 & \cite{mar13} \\
2008-2015 & $\gamma$-ray light curve from Fermi-LAT & $23 \pm 0.5$, $19.7 \pm 1.2$ $^{(e)}$ & n.c. $^{(e)}$ & \cite{bar15} \\
2008-2011 & $\gamma$-ray light curve from Fermi-LAT & n.a. $^{(e)}$ & $>6$ $^{(e)}$ & \cite{abd15} \\
2019 Jul & Lensing model from ALMA observations & 26--29 $^{(f)}$ & 1.07 $^{(d)}$ & \cite{mul20b}\\
2016 Mar-Sep & ALMA monitoring & $25 \pm 3$ $^{(g)}$ & $1.24 \pm 0.04$ $^{(g)}$ & This work \\
\hline
\end{tabular}
\tablefoot{ $(a)$ Most likely contaminated by contribution of the Einstein ring flux density;
$(b)$ from the determinant of the relative magnification matrix between the two vectors formed by the core-jet separation and offset between polarized and total-intensity emission for different epochs;
$(c)$ disputed by \cite{abd15}, as signal in the time-series analysis could come from the Moon's orbital motion;
$(d)$ fixed;
$(e)$ not constrained, if different $\gamma$-ray flares arise from different emitting locations, hence with potentially slightly different time delays and magnifications;
$(f)$ depending on the adopted value of $H_0$;
$(g)$ based on a parametric model of the continuum evolution (flux density and spectral index).}
\end{center} \end{table*}

The determination of the geometrical time delay between lensed images is interesting for measuring the Hubble constant, $H_0$, via the time-delay cosmography method \citep{ref64}. Now that the third lensed image of the quasar has been unambiguously detected by \cite{mul20b} ---thereby pinpointing the location of the lensing galaxy---,
the major remaining sources of uncertainty for the lens modeling of the \PKS1830\ system that can be constrained observationally are the time delay and the relative magnification factor between the NE and SW images.

Both quantities have been investigated with different methods and at different wavelengths (see Table~\ref{tab:timedelay}). At low radio frequencies, the required angular resolution of $<1\arcsec$ and the contamination from the Einstein ring make it difficult to disentangle the individual light curves from both images \citep{vanomm95,lov98}. As mentioned above, it is possible to measure the flux-density ratio at mm wavelengths using the foreground molecular absorption and the saturation level of high-opacity lines, where absorption almost completely masks the SW image and therefore makes it possible to retrieve the flux densities of both NE and SW images even with low-angular-resolution data \citep{wik01,mul06}. Time-series analyses of the $\gamma$-ray light curve benefit from long time ranges and daily monitoring\footnote{\PKS1830\ is monitored by the Fermi-LAT satellite on a daily basis: \url{https://fermi.gsfc.nasa.gov/ssc/data/access/lat/msl_lc/}}. However, results with this method are still controversial (e.g., \citealt{bar11,abd15}). Their interpretation might even be more complicated if individual $\gamma$-flares arise at different locations \citep{bar15}, implying slightly different time delays and magnification ratios. Overall, the results seem to converge toward a time delay in the range of $20-30$~days and a differential magnification of $\sim 1.0 - 1.5$, which seems consistent with a reasonable lens model \citep{mul20b}.

\begin{figure}[h] \begin{center}
\includegraphics[width=8.8cm]{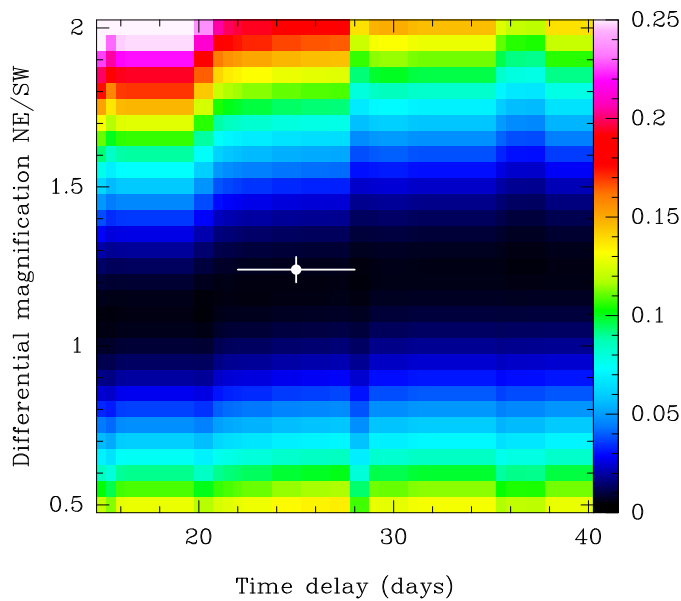}
\caption{Measure of randomness using the Von~Neumann estimator $\mathcal{E}_{\rm VN}$ (defined in Eq.\,\ref{eq:vonneumann}) as a function of the time delay and differential magnification parameters. The best-fit value obtained from a parametric model of the continuum evolution (Sec.\,\ref{sec:ContVarScenario}) is shown by the white mark.}
\label{fig:vonneumann}
\end{center} \end{figure}

The relatively featureless light curve during our ALMA 2016 monitoring and potential effects from milli- or micro-lensing prevent us from setting new strong constraints on the time delay and differential magnification of the \PKS1830\ system. Nonetheless, we applied a measure-of-randomness analysis (see, e.g., \citealt{che17}) using the Von~Neumann estimator ($\mathcal{E}_{\rm VN}$), which is defined as:
\begin{equation} \label{eq:vonneumann}
\mathcal{E}_{\rm VN}(\Delta t, \Rq) = \frac{1}{N-1} \sum _{i=1}^{N-1} \left [ \mathcal{F}(t_i) - \mathcal{F}(t_{i+1}) \right ]^2
,\end{equation}
\noindent where the datapoints $\mathcal{F}(t_i)$ correspond to the $(t_i; f_i^{\rm NE})$ flux-density measurements from the NE image augmented with the $(t_i-\Delta t; \Rq \times f_i^{\rm SW})$ points corresponding to the flux-density measurements of the SW image shifted in time by the time delay, $\Delta t$, and scaled by the differential magnification, \Rq, that is,
$\mathcal{F}(t_i) = f_{\rm NE}(t_i; f_i^{\rm NE}) \cup f_{\rm SW}(t_i-\Delta t; \Rq \times f_i^{\rm SW})$. By varying the two quantities $\Delta t$ and \Rq, we can calculate the corresponding values of $\mathcal{E}_{\rm VN}$. The assumption of minimum randomness (as in the $\chi^2$ minimization) points to the most likely values of the two parameters \Rq\ and $\Delta t$.
We show the results of this analysis in Fig.\,\ref{fig:vonneumann}. Given the quasi-monotonic and relatively featureless light curve of \PKS1830\ during our monitoring, there is a strong degeneracy between the two parameters. Without significant a priori information on the differential magnification, we cannot better constrain the time delay with only our 2016 flux-density measurements. In Sect.\,\ref{sec:ContVarScenario}, we further present a simple parametric model of the continuum evolution during the monitoring, from which we extract best-fit values of the time delay, namely $25 \pm 3$~days, and of the differential magnification, $1.25 \pm 0.04$, which fall in the valley of lowest Von~Neumann randomness in Fig.\ref{fig:vonneumann}.

\subsection{Differential polarimetry} \label{sec:Rpol}

To continue our analysis of the continuum emission properties of \PKS1830\ during our monitoring, we can also investigate the evolution of its polarization state. Although the ALMA data were not taken in full polarization mode, we can use the data obtained separately from the two orthogonal polarizers, XX and YY, in the alt-azimuth frame of the antennas to retrieve some source polarization properties using a differential polarimetry analysis (see, e.g., \citealt{mar15,mar16,mar19}). First, we can form the difference ratio $\Delta_{\rm XY} = (S_{\rm XX}-S_{\rm YY}$)/$S_{\rm I}$, where $S_i$ are the signal retrieved from the XX and YY polarizers, and I is the Stokes I total flux. We already see in Fig.\,\ref{fig:cont-monitoring}c that a clear polarization signal is present in the data, as $\Delta_{\rm XY}$ varies within $\pm 8$\% during the monitoring.

Furthermore, we can determine the flux-density ratios \RXX\ and \RYY\ from the XX and YY polarizers separately, and form the double polarization ratio, \Rpol:
\begin{equation}
\Rpol= \frac{1}{2} \left ( \frac{\RXX}{\RYY}-1 \right ),
\label{eq:rpol}
\end{equation}
\noindent as defined by \cite{mar15}. If the quasar is not polarized or if both lensed images have strictly the same polarization properties (polarization intensity and electric vector position angle, hereafter EVPA), then $\Rpol=0$. However, if  the two images have different polarization properties, which could be due to a faster polarization variability than the time delay, then \Rpol\ would be a nonzero quantity that would depend on the polarization intensity and EVPA angle differences between the two images as well as on the parallactic angle of the observations. With observations at different frequencies, it is furthermore possible to derive the Faraday rotation from \Rpol\ (see \citealt{mar15,mar16}).

The variations of \Rpol\  encode information on the polarization variability of the quasar. In Fig.\,\ref{fig:cont-monitoring}d, we see that \absRpol\  varies significantly between 0\%\ and $\sim 5$\% during the first half of the monitoring, but relaxes to $\sim 0$ after MJD~$=7580$. This suggests that the EVPA stabilized for a longer time interval than the time delay, after which the second image also stabilized. We note that the quasar fractional polarization eventually reaches a level of $\sim 5$\%, as shown by the nearly constant $\Delta_{\rm XY}$ after MJD~$=7580$. This value is comparable to other polarization measurements in 2015 (Band~9, \citealt{mar19}) and 2019 (Band~6, \citealt{mar20}), although the polarization variability can be significant over a shorter timescale than the time delay \citep{mar20}.

\subsection{Spectral index}

We finally show in Fig.\,\ref{fig:cont-monitoring}e the evolution of the spectral indices\footnote{We define the spectral index $\alpha$ as $S(\nu)=S_0 \left ( \frac{\nu}{\nu_0} \right ) ^{\alpha}$, $S(\nu)$ being the flux density as a function of the frequency.} of the two lensed images and their difference. The spectral indices were measured directly from the spectra corresponding to each lensed image after flagging channels associated with absorption features. We emphasize that any instrumental or calibration effects are removed when taking the spectral index difference $\Delta \alpha = \alpha_{\rm NE} - \alpha_{\rm SW}$ between the two lensed images.

Throughout the monitoring, the difference $\Delta \alpha$ was almost always positive, which reflects the rapid intrinsic variations of the quasar at a rate faster than the time delay, with the NE image having a steeper spectral index. Until the end of April 2016 (MJD $\sim 7500$), the spectral index is $\sim -0.7$. It then gets a little steeper, except for a sudden rise of the quasar spectral index for two measurements around MJD=7550 (15 and 22 June 2016), which corresponds to a small flux-density bump in the light curve. 

It is interesting to note that the spread of $\Delta \alpha$ (i.e., between $-$0.05 and 0.15) in the different visits of our six-month monitoring is comparable to the change of $\Delta \alpha$ measured by \cite{mar20} within a two-hour observation coincident in time with the onset of the 2019 record-breaking flare of \PKS1830. This suggests a considerable change of activity level between these two periods.

\subsection{Simple scenario of the continuum evolution} \label{sec:ContVarScenario}

The smooth and relatively slow flux-density evolution of the quasar suggests that the continuum spectra of the two lensed images should be similar, that is, that there are no clear features in the light curve on timescales short enough to introduce large differences between the NE and SW brightness or spectral index. Also, we do not consider the effects of micro- and milli-lensing here. If the decrease in source brightness was achromatic (i.e., if the flux density were to decay at the same rate for all frequencies), the spectral indices of the two images would be equal within the error bars. However, Fig.\,\ref{fig:cont-monitoring}e shows that the spectral indices of the two images are systematically different at almost all epochs, the spectral index of the NE image being about $\sim 0.05$ higher than that of SW. One natural way to explain this difference in spectral index is to have a slower brightness decrease at higher frequencies, such that the leading NE image always has a slightly flatter spectrum than the SW image.

Here we propose a simple parametric model of the blazar continuum emission able to simultaneously explain the monotonic flux-density decay and the (small) difference in spectral index between the NE and SW images. The model assumes that the source brightness consists in  two components: one that remains stable over time, that is, the core, and another that cools down adiabatically, that is, a moving feature, hereafter called a plasmon, injected in a conical jet. The total flux density in the time frame of the quasar $I(\nu, t)$ is then described as
\begin{equation}
 I(\nu, t) \propto I_c+I_p\left(\frac{t}{t_0}\right)^\beta\left(\frac{\nu}{\nu_0}\right)^{\alpha_p}, 
\label{eq:SimpleToyModel}
\end{equation}
\noindent where $I_c$ is the flux density of the core (assumed to have a flat spectrum), $I_p$ is the flux density of the plasmon at $t=t_0$ and $\nu=\nu_0$, $\alpha_p$ is the  spectral-index difference between the (optically thin) plasmon and the core, and $\beta$ accounts for the brightness decrease due to the adiabatic expansion of the plasmon. For sake of simplicity, we do not consider radiative losses, but only adiabatic expansion, meaning that $\alpha_p$ remains constant over time (i.e., time and frequency dependencies are decoupled).

If the plasmon propagates downstream of a conical jet at a constant speed, the $\beta$ index and $I_p$ are related to the initial distance between the plasmon and the jet base, as well as to the propagation speed, because (for adiabatic losses) the energy of each synchrotron-emitting electron will decrease as \citep{pac70}
\begin{equation}
\dot{E} \propto \frac{E}{(r-r_0)\sin{\theta}} = \frac{E}{v\sin{\theta}(t-t_0)},
\end{equation}
\noindent where $r$ and $v$ are the distance (to the jet base) and the propagation speed, respectively, $r_0$ is the initial distance, and $\theta$ is the jet opening angle. Integrating this equation for an initial electron population $N \propto E^{-p}$ (fixing $p=3$) and computing the resulting synchrotron spectrum (with the assumption that $r \gg r_0$) results in $\beta = \frac{1}{v\sin{\theta} + k}$, where $k$ depends on the choice of initial time, $t_0$. 

Using Eq.\,\ref{eq:SimpleToyModel}, we can estimate the spectral-index difference between the NE and SW images at any epoch:
\begin{equation}
    \Delta\alpha(t) = \alpha_{\rm NE} - \alpha_{\rm SW},
    \label{eq:SpectralIndexDiff}
\end{equation}
with 
\begin{equation}
\alpha_{\rm NE} = \frac{\log{I(\nu_2,t)}-\log{I(\nu_1,t)} }{\log{\nu_2}-\log{\nu_1}}
,\end{equation}
and 
\begin{equation} \label{eq:alphaSW}
\alpha_{\rm SW} = \frac{\log{I(\nu_2,t-\Delta t)} -\log{I(\nu_1,t-\Delta t)} }{\log{\nu_2}-\log{\nu_1}},
\end{equation}
\noindent where $\Delta t$ is the time delay between NE and SW images, $\nu_2$ and $\nu_1$ are the frequency pair used to compute the spectral index, and $t$ is the time for the NE image. The time at the observer's frame, $t^\prime$, would be stretched by the factor (which also affects $\Delta t$) given by
\begin{equation}
dt^\prime = (1+z)\frac{(v/c)\sin{\delta}}{1-(v/c)\cos{\delta}}dt, 
\end{equation}
\noindent where $z$ is the redshift of the quasar and $\delta$ is the angle between the jet and the line of sight. In our parametric model (Eq.\, \ref{eq:SimpleToyModel}), the values of $v$, $\delta$, and $\theta$ are all degenerated into $\beta$, $I_0$, and the choice of $t_0$. Hence, the time evolution of the model is only affected by $\beta$ and $\Delta t$ (in the observer's frame).

We show the results of this simple model simultaneously fitted to the light curve and spectral-index differences in Fig.\,\ref{fig:ToyModel}. The corner plot results of a Markov-chain Monte Carlo (MCMC) exploration of the plasmon injection time, $I_c$, $I_p$, $\beta$, $\alpha_p$, the time delay, $\Delta t$, and the differential magnification are shown in Fig.\,\ref{fig:model-corner}. In this MCMC exploration, we have added a 10\%  systematic uncertainty to the flux densities and spectral indices, and have computed the error function using the flux densities of NE and SW images, the spectral index of NE, and the difference in spectral index between NE and SW. Even though the model is not able to reproduce the scatter seen in spectral index, it is remarkable that such a simple model can capture the light-curve decrease and the (overall) spectral differences between the NE and SW images in a consistent way. The values of the fitting parameters obtained from the MCMC exploration are given in Table \ref{tab:ToyModelParameters}.

The constraint on the time delay, $\Delta t$, is highly model dependent (i.e., the choice of the time evolution in Eq.\,\ref{eq:SimpleToyModel}) and also changes if the relative weight between flux densities and spectral indices is changed in the error function. For instance, if the weight of the spectral index is increased or decreased by a factor 2, the value of $\Delta t$ decreases or increases by $\sim$ 3 or $7$ days, respectively. The fitted value of $\Delta t$ should therefore be taken with care. Still, the best fit value of $\Delta t = 25 \pm 3$~days is consistent with lensing models \citep{mul20b}. The differential magnification between the NE and SW images, on the other hand, is estimated to be 1.24 $\pm$ 0.04 and (together with the estimate of $\Delta t$) falls within the valley of lowest Von~Neumann randomness shown in Fig.\,\ref{fig:vonneumann}. The breaking of the time delay--differential magnification degeneracy that was noticed from the light-curve analysis is mainly due to the added information from the spectral index behavior, as $\Delta t$ is also present in Eq.\,\ref{eq:alphaSW}.

The main assumptions made for our modeling are as follows:

\begin{itemize}
\item We neglect effects from micro- and milli-lensing, and consider only intrinsic variations of the blazar.
\item The emission from the core, $I_c$, is stable in time and is considered to be optically thick ($\alpha_c = 0$).
\item A single event of plasmon injection.
\item A straight trajectory of the plasmon (i.e., a ballistic plasmon), so that there are no changes in the Doppler boosting factor due to changes in the viewing angle.
\item Radiative losses are negligible compared to adiabatic losses.
\item We fix the jet orientation and opening angle (the actual values of these quantities are degenerated into the fitting parameters).
\item The value of the core flux density is bound to be lower than the lowest flux density measured in the last epoch.
\end{itemize}

Our two-component model for the quasar continuum emission already provides a qualitative explanation for the evolution of the differential polarization, \Rpol. As the plasmon cools down as it propagates and expands in the jet, its contribution to the total flux density becomes smaller. Toward the end of our monitoring, the core (which would have a stable emission at timescales larger than the time delay) starts to dominate, and therefore produces similar polarization in the two images. Therefore, the differential polarization between the NE and SW images would vanish and \Rpol\ would tend to zero, as we indeed see in Fig.\,\ref{fig:cont-monitoring}d.

\begin{table*}[ht]
\caption{Best-fit values for the synchrotron-cooling ballistic plasmon model parameters (see Eq.\,\ref{eq:SimpleToyModel} and Fig.\,\ref{fig:model-corner}).
}
\label{tab:ToyModelParameters}
\begin{center} \begin{tabular}{llccc}
\hline
& Parameter  & & Value & Note \\
\hline
Lensing & Time delay (days)       & $\Delta t$ & 25 $\pm$ 3 & depends on plasmon model and data weights \\
& Differential magnification & \Rq & 1.24 $\pm$ 0.04 & weakly coupled to $\Delta t$ \\
\hline
Core & Flux density of the core (Jy) & $I_c$  & $0.178 \pm 0.010$ & bounded $<0.18$\,Jy \\
& Spectral index of the core & $\alpha_c$  & 0 & fixed \\
\hline
Plasmon & Age of the plasmon at $t_0$ (days) & & $120 \pm 30$ &  depends on plasmon model and data weights \\
& Flux density of the plasmon at $t_0$ (Jy) & $I_p$    & $1.51 \pm 0.09$ & weakly coupled to $t_0$ \\
& Power of the plasmon flux-density decay & $\beta$ & $-2.2 \pm 0.4$ & strongly coupled to $t_0$ \\
& Plasmon spectral index & $\alpha_p$ & $-0.89 \pm 0.02$ & \\
\hline
\end{tabular}
\end{center} \end{table*}

\subsection{Modeling with polarization evolution} \label{sec:ToyModelWithPola}

\begin{figure*}[h] \begin{center}
\includegraphics[width=\textwidth]{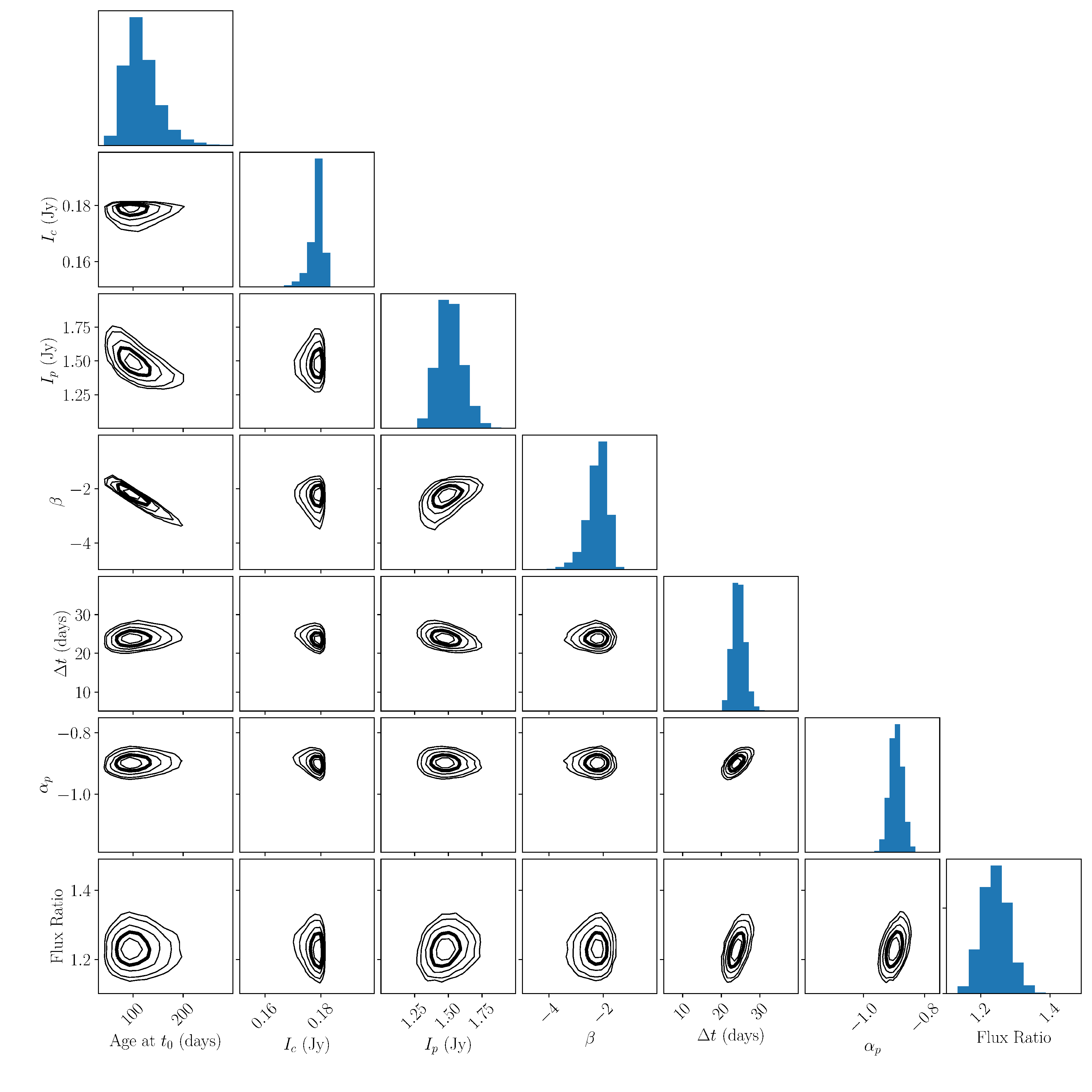}
\caption{Results from our plasmon modeling (Eq.\,\ref{eq:SimpleToyModel}) as obtained from MCMC (see also Table\,\ref{tab:ToyModelParameters}).}
\label{fig:model-corner}
\end{center} \end{figure*}

We propose an extension to the model given in Eq.\,\ref{eq:SimpleToyModel} to further account for the evolution of the polarization properties of \PKS1830\ during the monitoring. The model starts with the same properties as described in the previous section, that is, a stable core and an evolving plasmon, and adds parameters for the polarization properties. We assume that the core fractional polarization, $m_c$, evolves slowly, that is, over a much
longer  timescale than the time delay, and is therefore taken as a constant in the model. On the other hand, we allow the fractional polarization of the plasmon, $m_p$, to vary in time. Accordingly, in addition to Eq.\,\ref{eq:SimpleToyModel}, we describe the evolution of the Stokes $Q$ parameter (in the antenna frame) as:

\begin{equation} 
Q = m_cI_c\cos{(2\phi_c-2\psi)}
+ I_p\left(\frac{t}{t_0}\right)^{\beta}m_pF(t-t_0)\cos{(2\phi_p-2\psi)},
\label{eq:ModelWithPola} \end{equation}

\noindent where $\psi$ is the parallactic angle, $\phi_c$ and $\phi_p$ are the core and plasmon EVPAs, respectively, and $F(t)$ is a function that must decrease quickly with time to match the observed behavior of \Rpol\ toward the end of the monitoring without increasing to very high values early on in the life of the plasmon. All the required model observable quantities (i.e., $XX$/$YY$ and \Rpol) can then be computed from Stokes $Q$. By construction, the fit of the light curve (which uses total intensity) is decoupled from the polarization part, meaning that the parameters shown in Fig.\,\ref{fig:model-corner} will not change from the effects of Eq.\,\ref{eq:ModelWithPola}. 

\begin{figure}[h] \begin{center}
\includegraphics[width=8.8cm]{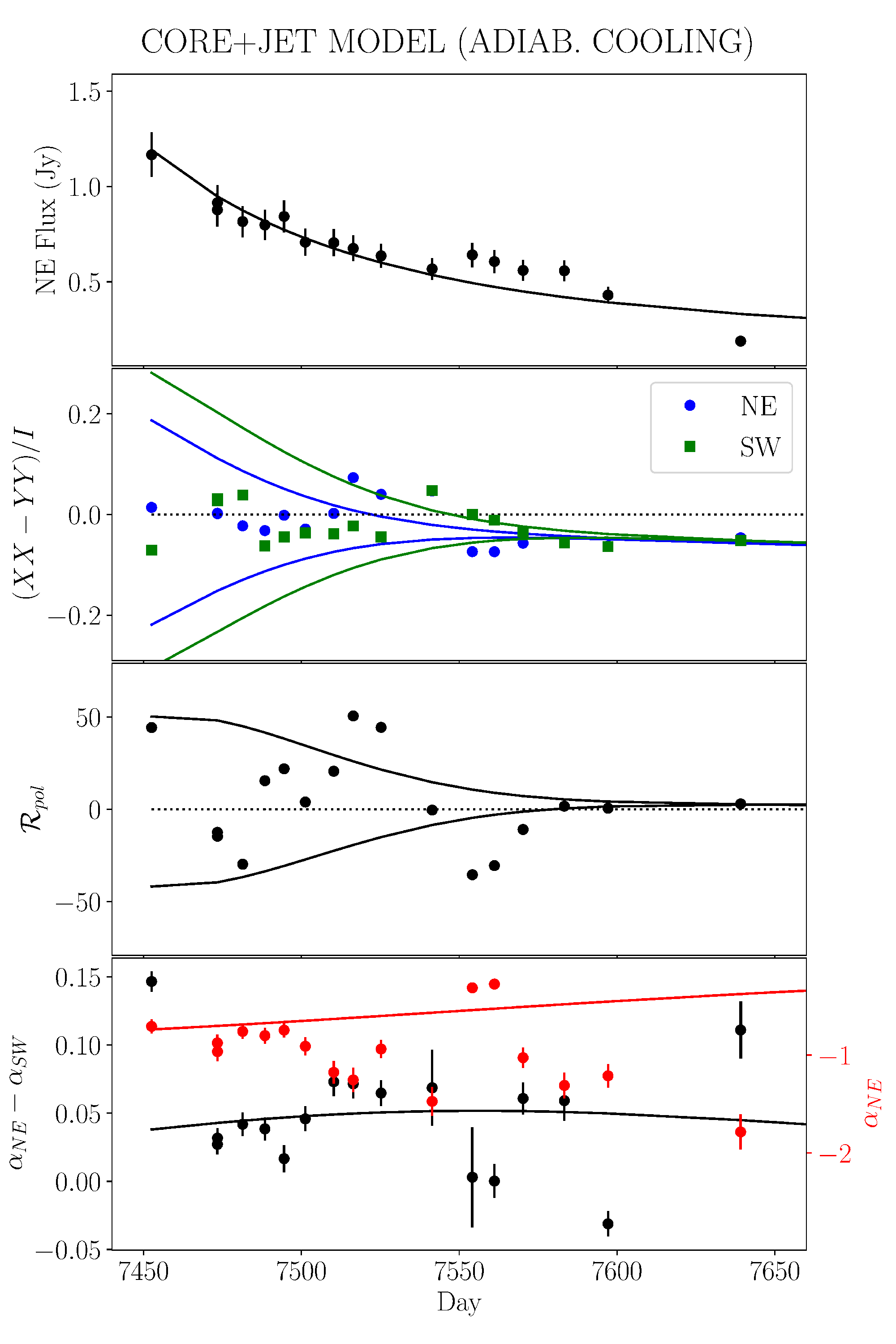}
\caption{Fit of the light curve, spectral index, and spectral-index difference between the NE and SW images using the parametric model provided by Eqs.\,\ref{eq:SimpleToyModel} and \ref{eq:SpectralIndexDiff}. 
The envelope of extreme model values for the differential polarization quantities is plotted against measurements using the model extension described in Sect.\,\ref{sec:ToyModelWithPola}.
The fit also provides the spectral index of the NE image (red points).}
\label{fig:ToyModel}
\end{center} \end{figure}

For the modeling of the data using Eq.\,\ref{eq:ModelWithPola}, we notice that the parallactic angle coverage of our data is very limited, which prevents us from putting robust constraints on the core and plasmon EVPAs. Hence, we can only test the model of Eq.\,\ref{eq:ModelWithPola} and try to recover a qualitative behavior of our differential polarization measurements. For the sole purpose of a qualitative exploration, we set $F(t-t_0)$ to be Gaussian, with a width of 100\,days, $m_c = 0.05$, and $m_p = 0.2$. The value of $m_c$ is taken to be equal to $(XX-YY)/I$ toward the end of the monitoring, where we assume that the contribution from the plasmon polarization becomes negligible. The result of this model is shown in Fig.\,\ref{fig:ToyModel} (second and third panels). The lines shown are envelopes that account for every possible value of $\phi_c$ and $\phi_p$, and assume that $\phi_c$ of the core is parallel to the parallactic angle in the last epochs of observation. From a qualitative point of view, the model recovers the basic behavior of converging toward a constant value of $(XX-YY)/I$ at late epochs (for both NE and SW), which in turn results in a null $\mathcal{R}_{pol}$ at these epochs. Unfortunately, it is not possible to constrain the model further, as (besides the limited parallactic-angle coverage of our observations) the EVPA of the plasmon may be changing at a faster rate than our time sampling.

\section{Results from spectral lines} \label{sec:linevar}

The dramatic drop in flux density of \PKS1830\ observed during our monitoring implies that there was substantial activity in the quasar, possibly with strong modifications of its continuum morphology. Only VLBI would be able to directly pinpoint such structural changes (see, e.g., \citealt{gar97}). Nevertheless, time variations of the absorption line profile from the foreground absorber due to a varying continuum illumination can also provide some insight into the quasar's activity and the absorbing screen properties.

Comparisons of a collection of HCO$^+$ J=2-1 \citep{mul08,mul14} and CS J=1-0 \citep{sch15} absorption spectra have already revealed drastic changes over  timescales of months and years. However, those observations were irregular and with consecutive visits spread over up to several months or even years.

Our dedicated 2016 monitoring with ALMA now allows us to track and investigate potentially subtle changes over a continuous six-month period and over timescales of down to $\sim 10$ days. We targeted the fundamental ground-state transitions of ortho-H$_2$O and CH (two $\Lambda$-doublets J=3/2--1/2), which can be observed simultaneously in one ALMA tuning in band~7. Those lines were observed by ALMA in 2012 \citep{mul14} and revisited in 2014 (Table~\ref{tab:ComplementarySpec}). The H$_2$O absorption is deeply saturated toward the SW image near $v \sim 0$~\kms\ \citep{mul14}, but its large opacity is well suited to investigating the variations of the line wings. The H$_2$O line is also well suited to probing changes in the presumably optically thin NE absorption. On the other hand, the CH absorption reaches an apparent peak opacity of $\tau \sim 1$ along the SW line of sight, complementing the H$_2$O line in tracing variations in the line center of the SW absorption.

The collection of absorption spectra of the H$_2$O ground-state transition and of the two CH $\Lambda$-doublets is shown in Fig.\,\ref{fig:monitoring-specSW} and \ref{fig:monitoring-specNE} for each visit of our 2016 monitoring for the SW and NE lines of sight, respectively. The rms noise values for each visit ---normalized to the continuum level--- are given in Table~\ref{tab:contdata-c3-ave} (last column) for the spectral window centered on the water line, which showed the lowest sensitivity. The two other spectral windows centered on the CH $\Lambda$-doublets show 10\%--20\% improved  sensitivity. The lowest rms noise levels, of namely $\sim 0.5$\% of the flux of the SW image (at the native channel resolution), were achieved at the start of the monitoring, when \PKS1830\ had a high flux density. The rms noise increased to $\sim 1$\% toward the end of the monitoring due to the quasar's drop in flux density. Only for one visit (2 June 2016) was the noise level up to $\sim 2$\% per channel due to poorer observing conditions. Hence, the ALMA monitoring provides us with the opportunity to trace variations of the absorption profile between two visits down to a level of about 1\%\ of the continuum level, at a velocity resolution of $<1$~\kms.

\subsection{Long-term variations (2012 - 2022)}

\begin{figure}[h] \begin{center}
\includegraphics[width=8.8cm]{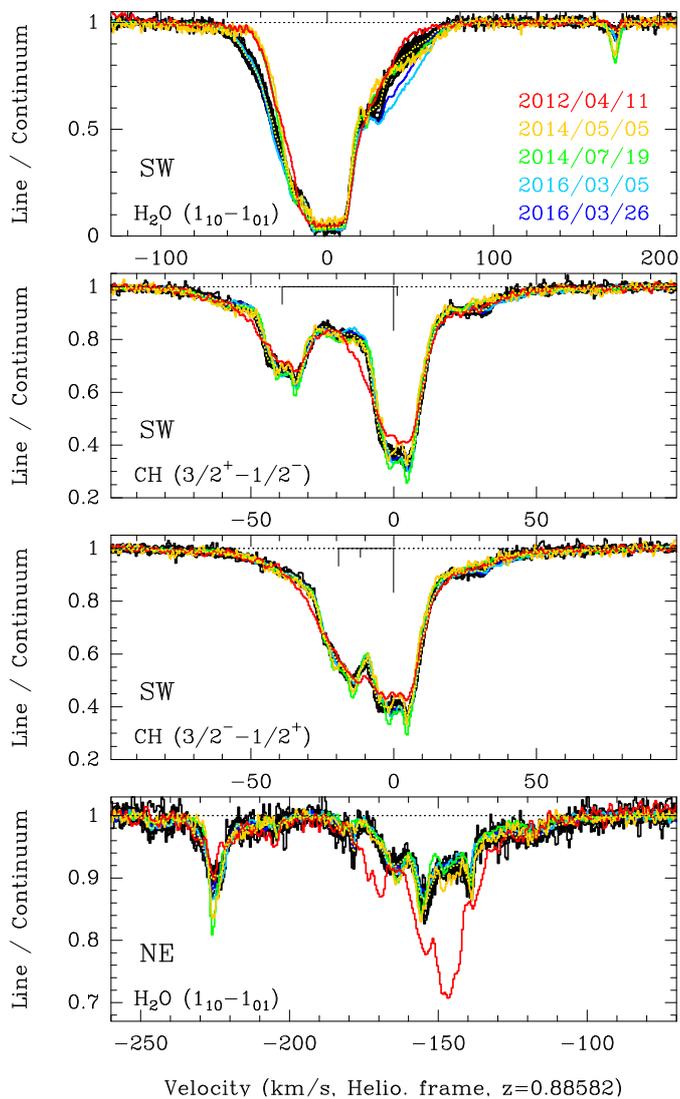}
\caption{Overlay of all ALMA spectra of H$_2$O (SW and NE lines of sight) and CH (SW only) observed toward \PKS1830\ between 2012 and 2016. Special profiles showing clear deviations are highlighted with the color code given in the upper box.}
\label{fig:var-long}
\end{center} \end{figure}

 Figure\,\ref{fig:var-long} shows all the available spectra of the H$_2$O (SW and NE lines of sight) and CH (SW line of sight only) lines taken with ALMA between 2012 and 2016 (observation details are listed in Tables~\ref{tab:contdata-c3-ave} and \ref{tab:ComplementarySpec}). The spectra with large deviations with respect to the majority are highlighted with a special color code.

\begin{figure}[h] \begin{center}
\includegraphics[width=8.8cm]{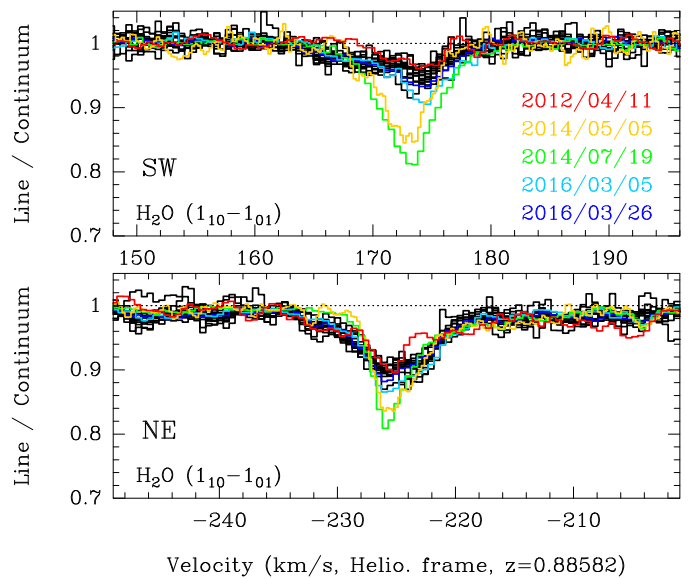}
\caption{Zoom onto the variations of the +174~\kms\ (top) and $-$225~\kms\ (bottom) features in the H$_2$O SW and NE spectra, respectively. Same as for Fig.\,\ref{fig:var-long}; special profiles showing clear deviations are highlighted with the color code given in the upper box.}
\label{fig:var-small}
\end{center} \end{figure}

The main variations for the H$_2$O spectra toward the SW image do not occur in the line center near $v \sim 0$~\kms, where the absorption is saturated, but in the wings of this component. The blue wing (i.e., $v \sim -20$ to $-60$~\kms) appears to have been stable between the 2012 and 2014 spectra, but jumped to a new level between 2014 and 2016. In contrast, the red wing (i.e., $v \sim +20$ to +60~\kms) shows more systematic variation, with the deepest absorption level being reached at the beginning of our survey in March 2016. This is not the case for the $v \sim +170$~\kms\ component, which was deepest in 2014 (Fig.\,\ref{fig:var-small}). At the same time, the CH profile of the SW absorption does not show significant variation, except for some changes between 2012 and later observations.

\begin{figure}[h] \begin{center}
\includegraphics[width=8.8cm]{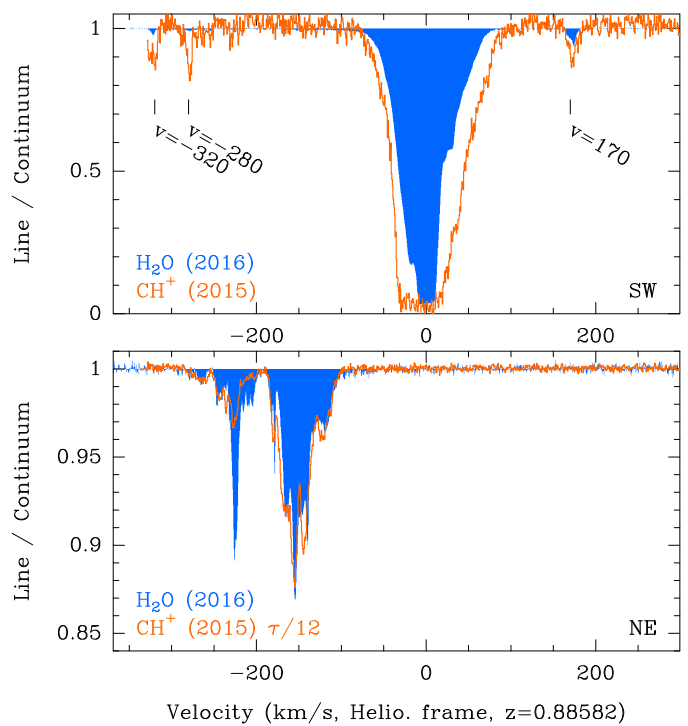}
\caption{Comparison of the average H$_2$O spectrum (filled blue) observed in 2016 with the ALMA CH$^+$ (1-0) spectrum (orange line) observed in 2015 \citep{mul17} for both lines of sight. For the NE spectrum, the opacity of the CH$^+$ line, assuming a filling factor of unity, is reduced by a factor 12.}
\label{fig:comparo-CH+}
\end{center} \end{figure}

Concerning the water absorption toward the NE image, there was a drastic decrease in the $v=-150$~\kms\ component between 2012 and 2014, whereas the subsequent spectra from 2014 to 2016 remain relatively unchanged. This major change is reminiscent of what happened between 2003 and 2006 in observations of the HCO$^+$ (2-1) line by \cite{mul08}, with the near disappearance of the same velocity component during this time interval. For the $v=-225$~\kms\ feature, unnoticed prior to 2009 ATCA observations by \cite{mul11}, the H$_2$O absorption varies continuously between 2012 and 2016 (Fig.\,\ref{fig:var-small}). To these water spectra, we can add the observations of the CH$^+$ (1-0) line in 2015 (reported by \citealt{mul17}). If we scale down the optical depth of CH$^+$ by a factor of $\sim 12$ (see Fig.\,\ref{fig:comparo-CH+}), its profile matches the $v=-150$~\kms\ feature  relatively well, but is not consistent for the $v=-225$~\kms\ feature, suggesting that one or the other significantly varied again between 2015 and 2016.

\begin{figure}[h] \begin{center}
\includegraphics[width=8.8cm]{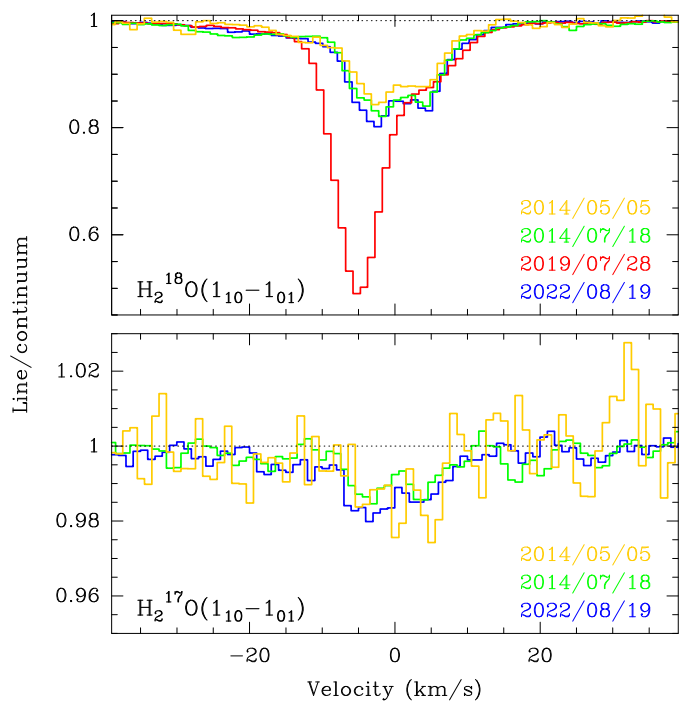}
\caption{Overlay of available ALMA spectra of H$_2^{18}$O and H$_2^{17}$O $1_{1,0}-1_{0,1}$ toward the SW image of \PKS1830.}
\label{fig:var-H218O}
\end{center} \end{figure}

Finally, there are a handful of observations of the rare water isotopologs H$_2^{18}$O and H$_2^{17}$O between 2014 and 2022 (Table~\ref{tab:ComplementarySpec} and shown in Fig.\,\ref{fig:var-H218O}). Those reveal a spectacular increase in the H$_2^{18}$O absorption for the $v \sim -5$~\kms\ component in the July 2019 spectrum, when \PKS1830\ was enduring a record-breaking $\gamma$- and radio flare (see Fig.\,\ref{fig:fluxratio-all}b, and, e.g., \citealt{mar20}). The integrated opacity of this component nearly increased by a factor four, while the absorption near $v=+5$~\kms\ did not change significantly. In particular, this remarkable change in absorption, together with the increase in the flux density of \PKS1830, allowed the detection of several deuterated species \citep{mul20a}. Interestingly, the most recent spectrum of H$_2^{18}$O obtained in August 2022 shows a return to almost the same profile as 2014, suggesting that the 2019 change was indeed due to an unusual event in \PKS1830, possibly related to a compact structure reminiscent of a typical Galactic dark cloud, as previously discussed by \cite{mul20a}.

\subsection{Time-variation statistics}

\begin{figure}[h] \begin{center}
\includegraphics[width=8.8cm]{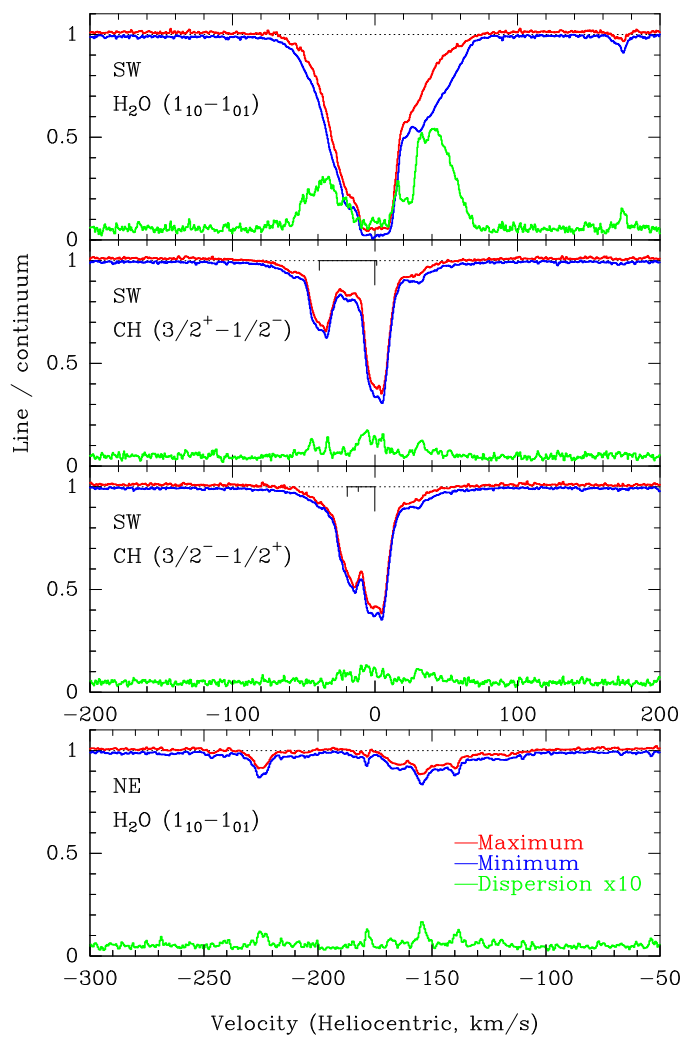}
\caption{Extrema and rms dispersion of the H$_2$O and CH absorption spectra during our 2016 monitoring (the CH profiles are not shown for the NE line of sight, because they contain only noise).}
\label{fig:stat-spec}
\end{center} \end{figure}

The extremes of the 2016 H$_2$O and CH absorption spectra are shown in Fig.\,\ref{fig:stat-spec}. The saturated region of the water absorption toward the SW image ($v \sim 0$~\kms) shows almost no variation, with the variation that we do see being due to a slight change of covering factor. It is interesting to note that the largest variations (up to $\sim 5$\% of the continuum level) occurred in the line wings, especially in the red wing ($v \sim +40$~\kms), but are not symmetric with respect to the line center. On the other hand, the variations seen in the CH spectra toward the SW image, of the H$_2$O $v \sim +170$~\kms\ component toward the SW image, and of the H$_2$O spectrum toward the NE image (i.e., all optically thin) are limited to an rms dispersion of $\lesssim 1$\%. 

\begin{figure}[h] \begin{center}
\includegraphics[width=8.8cm]{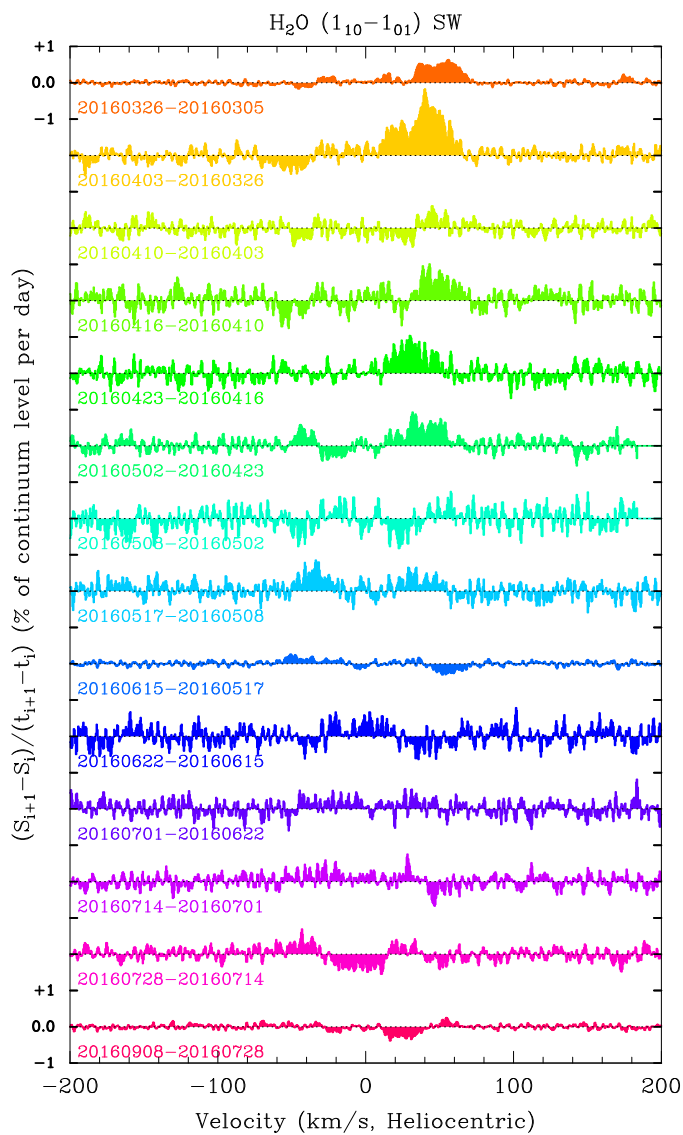}
\caption{Rate of the variations between consecutive observations of the H$_2$O absorption profile along the SW line of sight of \PKS1830.}
\label{fig:diff-rate}
\end{center} \end{figure}

Furthermore, we show the difference between consecutive spectra divided by their respective time separation (i.e., rate of change) in Fig.\,\ref{fig:diff-rate}. The strongest changes ---which are of up to $\sim 1$\% per day on average--- appeared at the beginning of the monitoring. Nevertheless, variations are not restricted to single time events, but apparently occurred continuously during the monitoring.

\subsection{Principal component analysis} \label{sec:pca}

\begin{figure}[h] \begin{center}
\includegraphics[width=8.8cm]{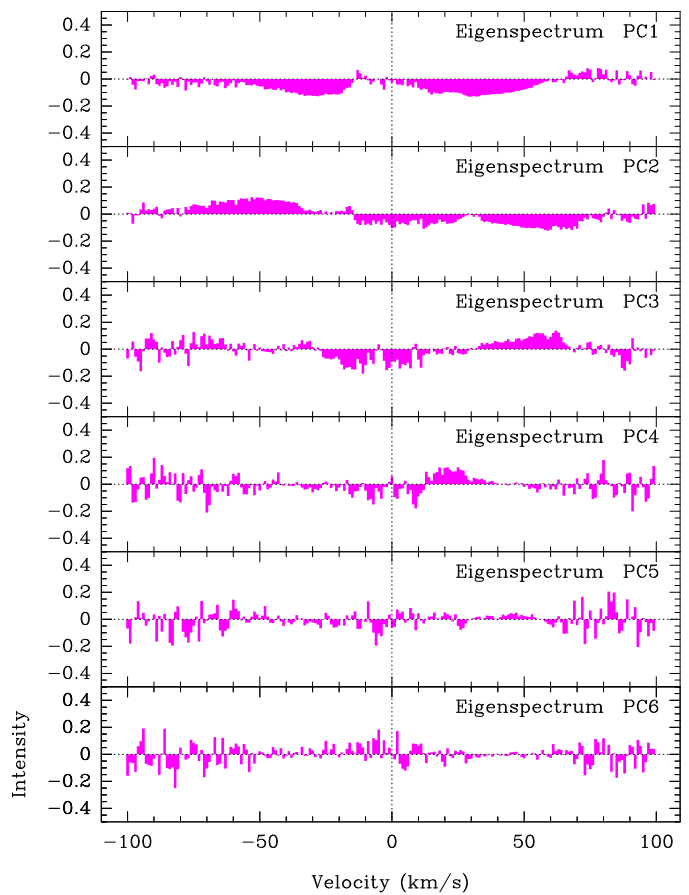}
\caption{Eigenspectra of the PCA decomposition of the H$_2$O spectra toward the SW image of \PKS1830.}
\label{fig:PCAEigenSpectra}
\end{center} \end{figure}

\begin{figure}[h] \begin{center}
\includegraphics[width=8.8cm]{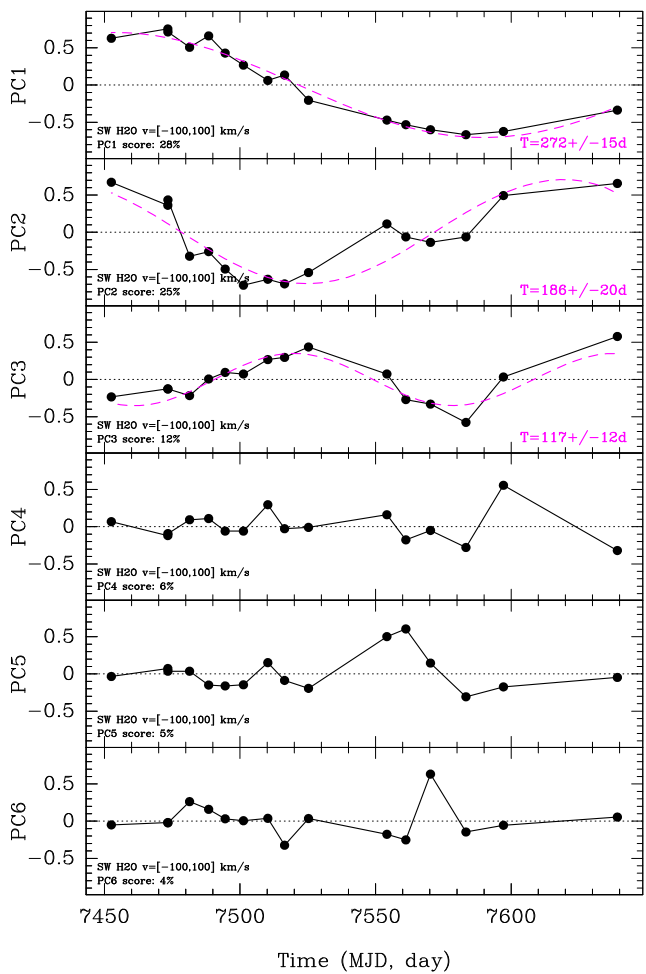}
\caption{Time evolution of the eigenvalues of the PCA decomposition of H$_2$O absorption spectra toward the SW image of \PKS1830.}
\label{fig:PCAEigenValues}
\end{center} \end{figure}

\begin{figure}[h] \begin{center}
\includegraphics[width=8.8cm]{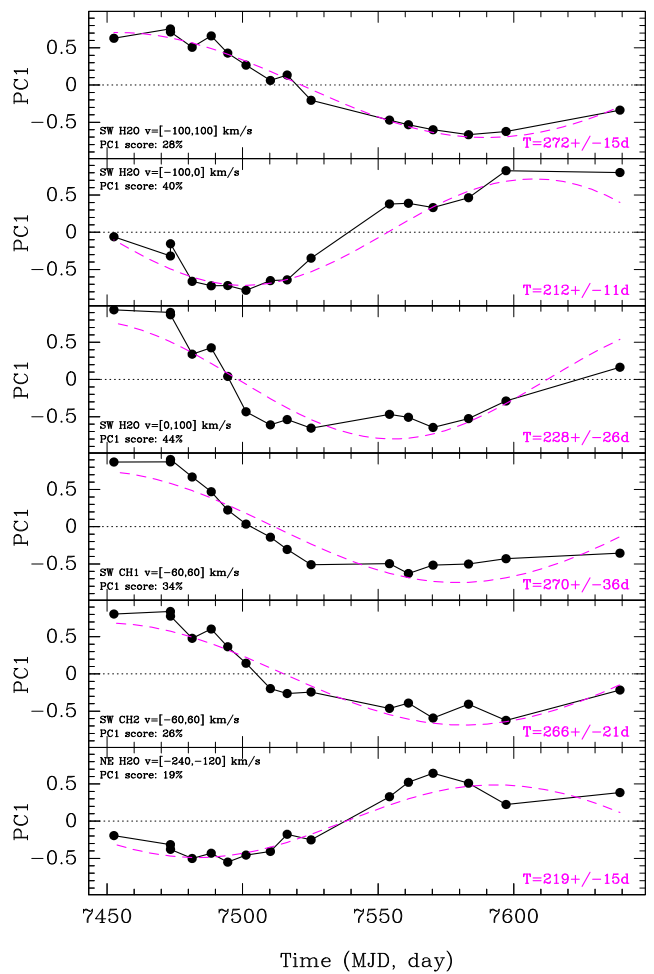}
\caption{Time evolution of the eigenvalues of the first component (PC1) of the PCA decomposition of various subsets of the monitoring spectra. The line of sight, species, considered velocity interval, and PC1 scores are given in each box.}
\label{fig:PC1all}
\end{center} \end{figure}

To extract more information from the variations of the absorption line profile, we ran a principal component analysis (PCA). This method searches for maximum variance through the data and outputs the results in new orthogonal axes, that is, principal components (hereafter PCA eigenspectra) with their associated (eigen)values in order of decreasing variance. We used the \texttt{sklearn.decomposition.PCA} class in the scikit-learn Python package \citep{ped11} to retrieve the eigenspectra (i.e., as a function of velocity) and eigenvalues (i.e., as a function of time) of the monitoring spectra. The results of the PCA decomposition for the H$_2$O spectra toward the SW image are shown in Figs.\,\ref{fig:PCAEigenSpectra} and \ref{fig:PCAEigenValues} for the eigenvectors and eigenvalues, respectively. The three most significant PCA components consist in spectral features in the blue and red wings of the absorption profile, which resemble to some extent the difference spectra shown in Fig.\,\ref{fig:diff-rate}. Most interestingly, the evolution of the three first eigenvalues (all together describing more than half the variance of the data) does not appear to be random but can be well described by a sinusoidal function across the entire duration of the monitoring.
Fitting such a sinusoidal function, we find periods of $272 \pm 15$\,days, $186 \pm 20$\,days, and $117 \pm 12$\,days, for PC1, PC2, and PC3, respectively. These periods are comparable to the duration of our monitoring and it remains to be seen if the evolution is truly periodic with monitoring over longer timescales. Using the spectra of the CH doublets along the SW line of sight instead of H$_2$O, or the H$_2$O spectra along the NE line of sight, we also find apparent wavy behaviors for PC1, as shown in Fig.\,\ref{fig:PC1all}. In addition, we separately checked for the velocity intervals corresponding to the blue and red wings of the H$_2$O absorption along the SW line of sight, and, again, find wavy signals. A final test, taking a velocity range outside of any absorption features (i.e., purely noisy channels), shows no significant structure in the PCA decomposition. All these measurements (Table~\ref{tab:PCAfits}; except overlapping velocity range for measurements from H$_2$O spectra) are independent, but still yield a consistent signal period of $\sim 230$~days for PC1, within error bars.

In order to test the PCA performances, we ran experiments synthetic spectra with similar noise properties and time sampling as in the ALMA monitoring, in which we introduced a time varying component (with a given period) in the form of a Gaussian perturbation with a given amplitude, full width at half maximum (FWHM), and velocity. We ran those synthetic spectra into the same PCA decomposition process as for the real observed spectra and the same fitting exercise with a sinusoidal function for the time evolution of the PC eigenvalues. We find that the first PC component PC1 succeeds in correctly  retrieving (i.e., within error bars of the fit) the period for perturbation signals with S/N down to a few percent. In other words, the PCA, and particularly its PC1, shows a good performance in retrieving this signal.

As it is difficult to imagine any periodic behavior in the absorbing clouds simultaneously
affecting both lines of sight and both H$_2$O and CH spectra, it is tempting to conclude that this wavy signal originates from the background illumination. This evolution could be connected to the model of a helical jet proposed by \cite{nai05}, for which the precession period was estimated to be of one year. This interpretation would indeed bring a natural explanation for the time variability of the absorption lines toward \PKS1830, but would need to be tested with new monitoring data over a longer time span. In this scenario, we would observe a slow modulation of the absorption due to the jet helicity, with occasional bursts or dips of absorption, depending on 
whether or not there is favorable illumination, whenever a new plasmon or a local flare occurs in the jet.

\begin{table}[ht]
\caption{Fit results for different PCA runs.}
\label{tab:PCAfits}
\begin{center} \begin{tabular}{cccccc}
\hline
Line of & Species & Velocity     &  PC1 score & $T_{\rm PC1}$ \\
sight   &         & range (\kms) &   (\%)     & (days)    \\
\hline    
SW & H$_2$O & [-100,100] & 28 & $272 \pm 15$ \\
   &     & [-100,0] & 40 & $212 \pm 11$ \\
   &     & [0,100] & 44 & $228 \pm 26$ \\
   & CH1 & [-60,60] & 34 & $270 \pm 36$ \\
   & CH2 & [-60,60] & 26 & $266 \pm 21$ \\
NE & H$_2$O & [-240,-120] & 19 & $219 \pm 15$ \\
\hline
\end{tabular}
\tablefoot{ $a)$ The errors quoted for the period are those coming out of the fit of a sine function to the PC1 eigenvalues versus time, all eigenvalues taken with the same weight.}
\end{center} \end{table}

\subsection{2016 averaged spectra}

\begin{figure}[ht] \begin{center}
\includegraphics[width=8.8cm]{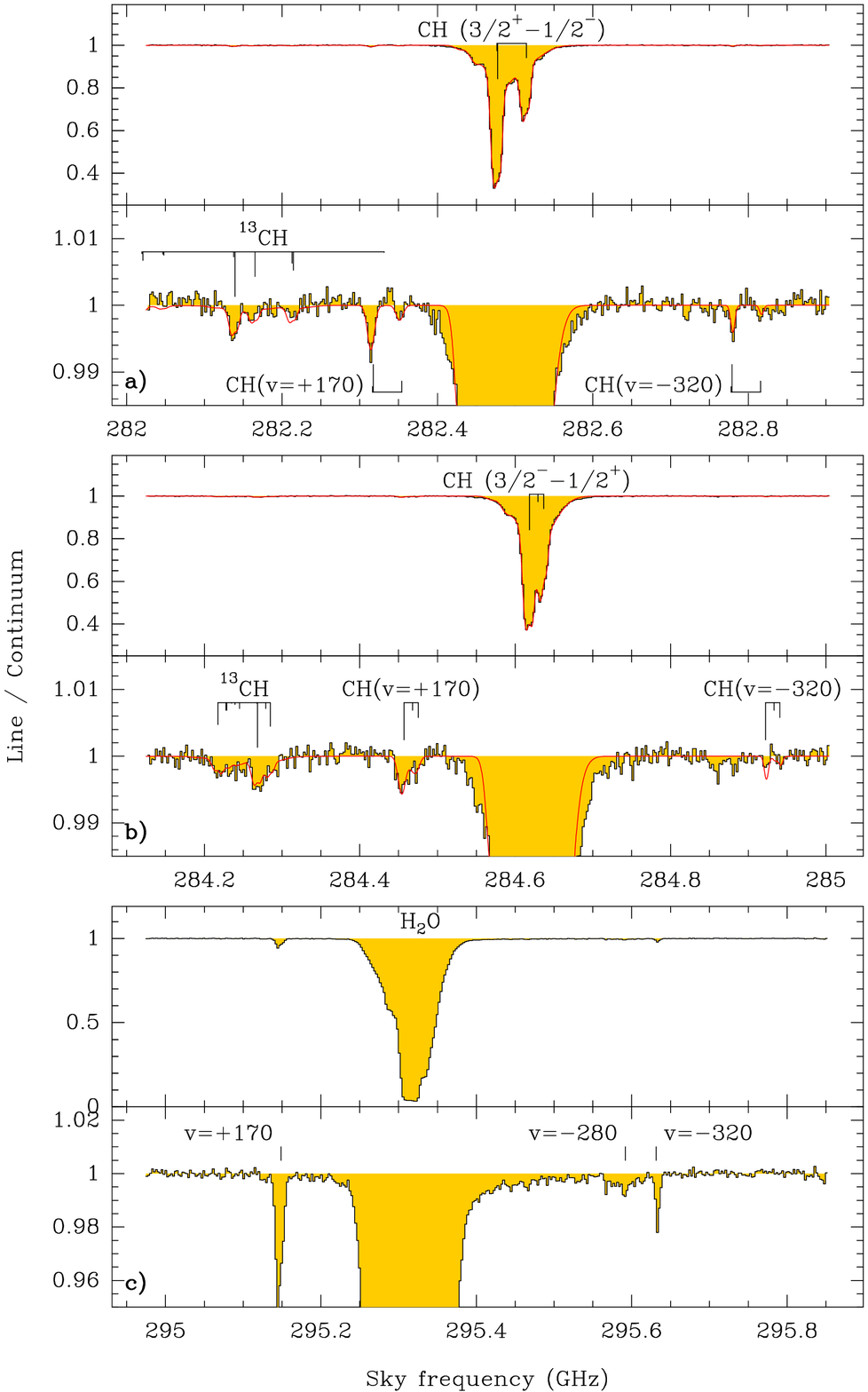}
\caption{Weighted-average spectra toward the SW image of \PKS1830, combining all visits in 2016. The hyperfine structure is shown for each CH $\Lambda$-doublet, with the strongest hfs component set at a velocity$v=0$~\kms, or as indicated otherwise. There are two boxes for each line, the first one showing the whole absorption spectrum and the second one showing a smaller range around the continuum level. Best-fit of the CH and $^{13}$CH absorption is shown in red.}
\label{fig:SuperSpecSW}
\end{center} \end{figure}

\begin{figure}[h] \begin{center}
\includegraphics[width=8.8cm]{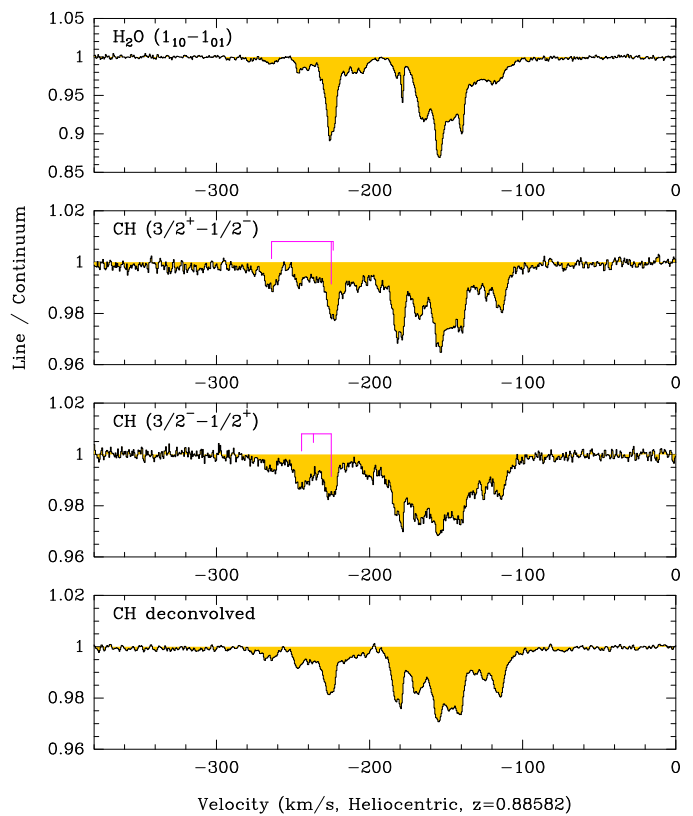}
\caption{Weighted-average spectra toward the NE image of \PKS1830, combining all visits in 2016. The hyperfine structure is shown for each CH $\Lambda$-doublet, with the strongest hfs component set at a velocity $v=-150$~\kms. The lower panel shows the CH hfs-deconvolved profile.}
\label{fig:SuperSpecNE}
\end{center} \end{figure}

We take advantage of the multiple visits of the 2016 monitoring to produce deep average spectra of H$_2$O and CH absorptions toward both images. To achieve the best sensitivity and quality, we removed a spectral baseline as a first-order polynomial (i.e., corresponding to the intrinsic spectral index of each lensed image) on each individual spectrum. We then averaged all spectra after weighting them individually by the square invert of the rms noise levels listed in Table~\ref{tab:contdata-c3-ave}. The resulting high-sensitivity spectra have a noise level of better than $0.1\%-0.2$\% of the normalized continuum level, as shown in Fig.\,\ref{fig:SuperSpecSW} and Fig.\,\ref{fig:SuperSpecNE}. These ``super'' spectra lead us to two discoveries along the SW line of sight: the detection of the weak absorption from the $^{13}$CH isotopolog and a remarkable absorption trough of 500~\kms\ in width, both of which are discussed in the following subsections.

\subsubsection{Detection of $^{13}$CH} \label{sec:detection13ch}

The rare isotopolog $^{13}$CH was recently detected for the first time in the ISM of the Milky Way by \cite{jac20}. As CH is ubiquitous in the ISM, with optically thin lines, and is expected not to be affected by fractionation in its chemical formation, it appears to be an excellent species for measuring the $^{12}$C/$^{13}$C elemental isotopic ratio (see a detailed discussion by \citealt{jac20}). This isotopic ratio is an interesting probe of stellar nucleosynthesis history. Chemical-evolution models (e.g., \citealt{pra96,kob11}) predict a positive $^{12}$C/$^{13}$C gradient in the Milky Way, increasing with galactocentric distance and decreasing with time. 


Here, we present the first detection of $^{13}$CH toward the SW image of \PKS1830\ in the average spectra of the 2016 monitoring data. The detection is robust, with its typical fine structure (see Table~\ref{tab:spectro}) present in both $\Lambda$-doublets, as shown in Fig.\,\ref{fig:SuperSpecSW}a and b. Nevertheless, the $^{13}$CH absorption signal is very weak, partly due to the optical thinness of CH, to the large $^{12}$C/$^{13}$C isotopic ratio, and to the dilution of the signal among the many hyperfine components. We were able to obtain a detection in absorption toward \PKS1830\ thanks to the brightness of the quasar and the high sensitivity of ALMA, providing a very high signal-to-noise ratio after averaging all the monitoring spectra.

In order to estimate the $^{12}$CH/$^{13}$CH ratio, we performed a simultaneous fit of the two $\Lambda$-doublets of $^{12}$CH and $^{13}$CH for the SW absorption spectrum using the same intrinsic velocity profile for both CH and $^{13}$CH convolved to their corresponding fine structure, respectively, and with the $^{12}$CH/$^{13}$CH ratio as free scaling parameter. We made several attempts with increasing numbers of Gaussian velocity components to reproduce the absorption profile of CH, and obtained satisfying results using about ten components.
During the process, we found that the ratio was relatively stable around approximately $150$, while the uncertainties and fit residuals decrease with an increasing number of velocity components. Our final $^{12}$CH/$^{13}$CH ratio is $150 \pm 10$. The physical relevance of every single velocity component in the fit is not certain, but it is clear that a mix of narrow and broad components is needed to reproduce the profile, in particular with some large components for the wings of the main $v=0$~\kms\ absorption feature.

Compared to the $^{12}$C-/$^{13}$C- isotopolog ratios from other species in the SW line of sight toward \PKS1830, the $^{12}$CH/$^{13}$CH ratio is the highest measured so far: $97 \pm 6$ for $^{12}$CH$^+$/$^{13}$CH$^+$ \citep{mul17}, $62 \pm 3$ for CH$_3$OH/$^{13}$CH$_3$OH \citep{mul21}, and even lower values $20-50$ for HCO$^+$, HCN, and HNC \cite{mul06,mul11}. This strongly suggests that all these other species may be affected by fractionation issues. A detailed analysis of the $^{12}$C-/$^{13}$C- isotopolog ratios is beyond the scope of this study and will be presented elsewhere.

A $^{12}$C/$^{13}$C ratio $\sim 150$ in the $z=0.89$ absorber is higher than the values observed within the inner 16~kpc of the Milky Way (see, e.g., the measurements of $^{12}$C/$^{13}$C with galactocentric distance in Fig.\,4 by \citealt{jac20}); as such, it probably reflects a true evolution effect between the $z=0.89$ absorber and the Milky Way today, as also found for other isotopic ratios (e.g., from N, O, S, Si, \citealt{mul06,mul11}).
 

\subsubsection{Weak and broad water absorption trough in the SW sightline} \label{sec:watertrough}

In addition to the long-known main absorption features at $v\sim 0$~\kms\ and $v\sim -150$~\kms\ toward the SW and NE images, respectively \citep{wik98, mul06}, \cite{mul11} discovered narrow velocity components at velocities of $v=-300$, $-224$, and +170~\kms\ in an unbiased 7~mm absorption survey with the Australian Telescope Compact Array. The $-224$ and +170~\kms\ components were later identified in ALMA data as occurring toward the NE and SW image, respectively \citep{mul14}, but the  $v=-300$~\kms\ component was not confirmed. However, two narrow velocity components at $v=-320$ and $-280$~\kms\ were later observed in 2015 in the CH$^+$ (1-0) ALMA spectrum toward the SW image (Fig.\,\ref{fig:comparo-CH+} and \citealt{mul17}), although near the edge of the spectral window. Both components are now readily confirmed in the H$_2$O ``super'' spectrum toward the SW image, as can be seen in Fig.\,\ref{fig:SuperSpecSW}c.

The origin of these components with large velocity spans in the line of sight is unclear, and they could be due to a high inclination of the absorber, large velocity gradients in the disk, or an extra-planar gas component, such as high-velocity clouds in the halo as observed around the Milky Way (as also discussed previously by \citealt{mul14}). They are not explained in the absorption kinematic model by \cite{com21}, which reproduces the H\,I and OH absorption spectra seen at cm wavelengths and the molecular absorption at mm wavelengths.

In addition, the ALMA ``super'' spectrum of the H$_2$O absorption now reveals an even more surprising structure in the form of a weak ($\tau < 0.01$) continuous absorption trough covering a wide velocity range of nearly 500~\kms\ toward the SW image (Fig.\,\ref{fig:SuperSpecSW}c). This is much wider than the continuous absorption span of $\sim 200$~\kms\ along the NE line of sight. It could be that this absorption trough reveals the presence of circumgalactic gas (e.g., \citealt{bah69, tum17}), although with up to such high velocities, it is not clear whether or not the gas would remain tied to the galaxy, depending on the halo-escape velocity, or would then (literally) form a "water fountain". Given the lack of constraints on the distribution of such material in the halo (probably fragmented in filamentary structures), it is not possible to get a meaningful estimate of its mass content. In any case, we again emphasize the weakness of this signal, which would be extremely difficult to trace in other galaxies unless a background continuum source as bright as \PKS1830\ were present.

\subsection{Water isotopologs and oxygen isotopic ratios}

With different water isotopologs observed close in time in the 2014 and 2022 datasets (Table~\ref{tab:ComplementarySpec}), we can derive the H$_2^{16}$O:H$_2^{18}$O:H$_2^{17}$O isotopolog ratios of the water family. Those are not expected to be affected by fractionation effects and should therefore reflect the true elemental isotopic ratios of oxygen. For a given epoch, we assumed the same opacity profile (starting with five Gaussian velocity components) for all water isotopologs, as well as for CH when available in order to make the link between the high-opacity line of H$_2^{16}$O and the optically thin rare water isotopologs. We then fitted the opacity ratios as free scaling factors. When including the main water isotopolog, we also added the continuum-covering factor as a free parameter of the fit; otherwise, we fixed it to unity for the fit of the H$_2^{18}$O and H$_2^{17}$O isotopologs alone. The results of our fitting exercises are summarized in Table~\ref{tab:FitOxygenRatios}. Due to the somewhat better signal-to-noise ratio of the data in July 2014, we used a higher number of Gaussian velocity components to obtain a better match of the profile and decrease the reduced $\chi$-squared. However, we found that the fitted ratios were not drastically affected and remained similar within the uncertainties. The values of the $^{16}$O/$^{18}$O ($= 65.3 \pm 0.7$ after combining the measurements with their statistical weights) and $^{18}$O/$^{17}$O ($= 11.5 \pm 0.5$) ratios are consistent with the previous measurements made from the HCO$^+$ isotopologs by \cite{mul06,mul11}, albeit with much better accuracy. Regarding the $^{12}$C/$^{13}$C ratio discussed above (Sect.\,\ref{sec:detection13ch}), the oxygen isotopic ratios point to a different ISM isotopic composition from that of the solar neighborhood in the Milky Way today (e.g., \citealt{luc98}), namely one that is less processed and most likely dominated by massive-star nucleosynthesis (see also, e.g., \citealt{wal16,mart19,tan19} and references therein).

\begin{table*}[ht]
\caption{Results for a global fit using the same opacity profile for different combinations of lines and epochs.}
\label{tab:FitOxygenRatios}
\begin{center} \begin{tabular}{cccccccc}
\hline
Species & Date & $N_{\rm gauss}$ $^{(a)}$ & \fc $^{(b)}$ & H$_2$O/CH $^{(c)}$ & H$_2^{16}$O/H$_2^{18}$O $^{(c)}$ & H$_2^{18}$O/H$_2^{17}$O $^{(c)}$ & $\chi^2_r$ $^{(d)}$\\ 
\hline
H$_2$O, H$_2^{18}$O, H$_2^{17}$O, CH & 2014/05/05 & 5 & 0.942(3) &  $8.6 \pm 0.1$  & $68.5 \pm 1.4$ &  $9.4 \pm 1.3$ & 1.09 \\
H$_2$O, H$_2^{18}$O, H$_2^{17}$O, CH & 2014/07/18-19 & 5 & 0.975(2) & $9.18 \pm 0.07$ & $65.9 \pm 1.0$ & $14.0 \pm 2.3$ & 3.85 \\
H$_2$O, H$_2^{18}$O, H$_2^{17}$O, CH & 2014/07/18-19 & 9 & 0.975(2) & $8.90 \pm 0.05$ & $64.2 \pm 0.8$ & $14.1 \pm 1.8$ & 2.29 \\
H$_2^{18}$O, H$_2^{17}$O  & 2014/05/05 & 3 & 1 $^{(e)}$ & -- & -- & $9.6 \pm 2.1$ & 1.35 \\
H$_2^{18}$O, H$_2^{17}$O  & 2014/07/18 & 3 & 1 $^{(e)}$ & -- & -- & $13.9 \pm 1.1$ & 1.31 \\
H$_2^{18}$O, H$_2^{17}$O  & 2022/08/19 & 3 & 1 $^{(e)}$ & -- & -- & $11.0 \pm 0.6$ & 1.25 \\
\hline
\end{tabular}
\tablefoot{ $a)$ Number of Gaussian components in the absorption profile; $b)$ Continuum filling factor; $c)$ Line opacity ratio (for CH, this refers to an hfs component of relative strength unity); $d)$ Reduced chi-squared; $e)$ Fixed.}
\end{center} \end{table*}


\section{Discussion} \label{sec:discussion}

\subsection{Analysis of the absorption profiles and their time variations} \label{sec:linevaranalysis}

The absorption spectrum $S_{\nu}$ resulting from an absorbing screen with opacity $\tau_{\nu}(x,y)$  ($x$ and $y$ being in the plane of the sky) in front of a background continuum illumination $i(x,y)$ and normalized to it is
\begin{equation} \label{eq:Inu}
 S_{\nu} = \frac{1}{I_{tot}} \iint i(x,y){\rm e}^{-\tau_{\nu}(x,y)}dx dy,
\end{equation}
\noindent where $I_{tot}= \iint i(x,y) dx dy$ is the total continuum intensity. Therefore, the normalized absorption spectrum is the continuum intensity-weighted average of the exponential of the opacity distribution of the illuminated material: $S = \langle {\rm e}^{-\tau(x,y)} \rangle_i$. In the case where the continuum emission is spatially extended but unresolved by the observations, the distributions i(x,y) and $\tau(x,y)$ are unknown, and we can only measure $I_{tot}$ and the resulting normalized absorption spectrum $S_{\nu}$. If we consider a representative average opacity $\langle \tau'_{\nu} \rangle$, Eq.\,\ref{eq:Inu} can be expressed as:
\begin{equation} \label{eq:Inu2}
S_{\nu} = 1-\fc \times (1-{\rm e}^{- \langle \tau'_{\nu} \rangle}),
\end{equation}
where we introduce the source-covering factor \fc, which is defined as
\begin{equation} \label{eq:fc}
 f_c = \frac{\iint_{\Omega_{abs}} i(x,y) dx dy}{I_{tot}},
\end{equation}
\noindent where $\Omega_{abs}$ is the solid angle fraction of the background continuum actually covered by absorbing material (weighted by intensity); by construction: $0 \le f_c \le 1$.

\begin{figure}[h] \begin{center}
\includegraphics[width=8.8cm]{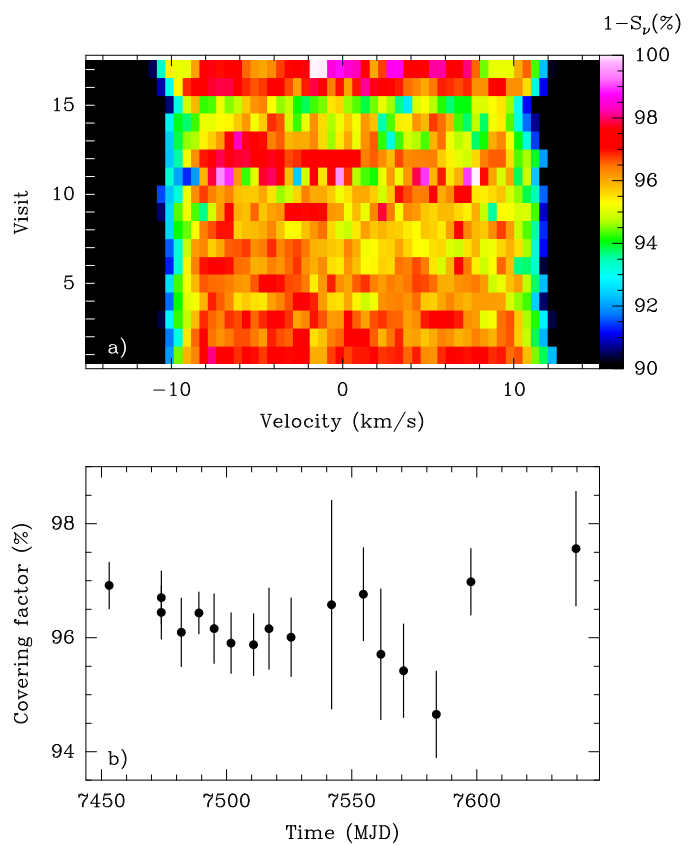}
\caption{Measurements of the covering factor across the different visits of our 2016 monitoring. {\em (a)} Saturation level ($1-S_{\nu}$) at the center of the H$_2$O line for all the visits in 2016. {\em (b)} The values of all spectral channels with velocities $|v|< 8$~\kms\ are averaged to provide measurements of the continuum-covering factor as a function of time.}
\label{fig:fc}
\end{center} \end{figure}

The most straightforward way to measure the covering factor is to use the saturated region of a very optically thick line, such as the H$_2$O $1_{10}$-$1_{01}$ in our case ($\tau \gtrsim 10$).  Figure \,\ref{fig:fc} shows the measurements of \fc\ across the different visits of our 2016 monitoring. The covering factor is $\sim 96$\% with only mild variations of the order of 1\%, which contrast with the large variations of the flux density between the beginning and the end of our monitoring (Fig.\,\ref{fig:cont-monitoring}). The values of \fc\ are consistent with previous measurements of saturated lines \citep{mul14}.

To further characterize the absorption variations, we can assume that the absorbing screen is constant in time, that is, $d\tau(x,y)/dt =0$.
This is justified by the fact that the timescales associated with potential causes of variations in the absorber are much longer than those we are interested in, even for small-scale structures on AU- to pc scales. For example, a structure of 10~AU with a velocity of 200~\kms\ in the plane of the sky (e.g., due to rotation for a face-on spiral galaxy) would have a crossing time in front of a fixed point-like continuum of three months. For a cloud collapse, the free-fall timescale is $t_{ff} \sim {\rm 2 months} \sqrt{ \frac{(R/AU)^3}{(M/M_{\odot})}}$, where $R$ is the cloud size and $M$ its mass. Finally, we can also exclude variations due to a change of the chemical composition of the absorbing material, because typical chemical timescales are orders of magnitude longer than years (e.g., \citealt{tie05,val17}).
Therefore, we need only consider time variations due to structural changes of the quasar's morphology, that is, change of $i(x,y)$ with time.

If the absorbing screen is uniform, where $\tau(x,y)=\tau_0$, there can be no variation of the normalized absorption spectrum, regardless of variations of $i(x,y)$. Consequently, the mere observation of variations of the absorption spectra implies that there are opacity structures in the absorbing screen at scales smaller than the continuum size. Similarly, a simple change of $i(x,y)$ by a multiplicative factor would not induce variations of the absorption spectrum due to the normalization by $I_{tot}$, and therefore variations of $S_\nu$ also imply small-scale structures in the quasar's continuum emission. Hence, the absorption variations reflect the coupling of small-scale structures between the continuum $i(x,y)$ and absorbing screen $\tau(x,y)$.

Taking Eq.\,\ref{eq:Inu2} as a simplified description of the absorption spectra and assuming that all continuum variations result in a change of the average opacity, then the derivative d$S_{\nu}$/d$\tau$ yields an estimate of the variation; we get
\begin{equation} \label{eq:varS}
\Delta S \propto \tau {\rm e}^{-\tau} \left ( \frac{\Delta \tau}{\tau} \right ).
\end{equation}
\noindent The function $x {\rm e}^{-x}$ peaks sharply for $x=1$ and tends rapidly to zero for large $x$, explaining why the maximum variations of the water absorption are seen in the line wings but are strongly attenuated at the saturated line center.

\subsubsection{Possible causes of time variations of the continuum distribution}

In the long term (months to years), we have seen some drastic changes in the absorption profiles, such as the near disappearance of the NE absorption near $-150$~\kms\ between 2003 and 2006 \citep{mul08}, its drastic decrease between 2012 and 2014, or the doubling of the absorption depth for the $v=-5$~\kms\ velocity component in the H$_2^{18}$O SW profile observed in 2019 (Fig.\,\ref{fig:var-H218O} and \cite{mul21}). This last event was occurring at the time of a record-breaking flare of \PKS1830\ at $\gamma$-rays and radio. It is possible that one or another cloud becomes occasionally illuminated (or highlighted) by a new continuum component. This "occasional event" scenario is consistent with the fact that the last H$_2^{18}$O spectrum in 2022 shows that the absorption returned to almost the same profile as pre-2019, which we could then qualify as a "quiescent absorption profile". Nevertheless, at this stage, we cannot exclude the possible effect of milli-lensing on timescales of months to years. In 2019, the flux-density ratio was indeed in the low range of values (i.e., either the NE image was fainter than usual with respect to the SW image, or the SW image was brighter than the NE image). We would need more statistics of drastic absorption-change events to investigate possible a connection with $\gamma$-ray and radio flares. 

On shorter timescales of weeks to months, the effect of micro-lensing on the variability of absorption spectra was investigated by \cite{lew03}. The picture is highly complicated by the large parameter space defined by absorbing cloud properties (shape, size, absorption profile) and distribution with respect to the background continuum illumination (all unknown), but \cite{lew03} validate the fact that micro-lensing can potentially induce significant modulation of the absorbing signal provided absorbing clouds have subparsec-scale structures.

An object of mass $M$ in the lens plane can produce significant variability by (micro-)lensing for a typical source size below the Einstein radius:
\begin{equation} \label{eq:thetaE}
\theta_{\rm E} = \sqrt{\frac{4GM}{c^2}\frac{D_{ls}}{D_{ol}D_{os}}}
,\end{equation}
\noindent where $D_{ij}$ are the angular diameter distances between the observer $o$, the lens $l$, and the source $s$, respectively. For \PKS1830\ and a stellar-mass object, $\theta_{\rm E} \sim 2$~$\mu$as.

The size of the SW core was measured to be $\theta_{\rm SW} \sim 0.1$~mas at $\lambda \sim 1$~cm and to scale with a $\sim \lambda^2$ dependence at radio--cm wavelengths \citep{gui99}. Extrapolating this behavior to mm wavelengths, we would obtain a size of 1~$\mu$as at $\lambda \sim 1$~mm. However, broadening  due to interstellar scattering may have a far less significant effect at mm/submm wavelengths. In fact, \cite{mul21} estimated a flatter dependence in $\lambda^{0.3}$ from the covering factors derived from a set of submm methanol lines. Extrapolating the size measurement at 1~cm with this flatter dependence, we obtain a size estimate of about 50~$\mu$as at 1~mm\footnote{This corresponds to 0.4~pc in the $z=0.89$ absorber.}, which is roughly one order of magnitude larger than $\theta_{\rm E}$. Therefore, it is possible that micro-lensing induces variability on timescales of weeks to months, which corresponds to the caustic crossing time \citep{lew03}.

On the other hand, the scenario of a helical jet would naturally induce variations of the continuum illumination with a clear periodicity. In turn, this periodicity may be imprinted in the variations of the absorption profile, which could be what we observe in the 2016 spectra, as suggested by our PC analysis (Sect.\,\ref{sec:pca}). In Sect.\,\ref{sec:ContVarScenario}, we propose a simple parametric model of a single ballistic plasmon in the jet to account for the continuum evolution during the 2016 monitoring. The ballistic assumption would not necessarily be in contradiction with the helical jet scenario. The toy model would also work even if the plasmon path is helical within the jet and its opening angle. There could be a time-changing Doppler boosting that would be added to the power-law decay of the toy model (depending on how open the helical path is and how close the jet line is to the line of sight), but the effect could be hidden below the flux-density uncertainties. Moreover, the facts that {\em (i)} we see a clear sinusoidal signal in the PCA and that {\em (ii)} the largest variations of the absorption profile happened at the beginning of the monitoring may support the single plasmon injection event. Indeed, supposing there were more plasmons at different points of the helical jet, one would see a "dirty" superposition of sinusoids with different phases. Therefore, having one simple sinusoidal signal in the absorption variation may be telling us that the illumination pattern of the absorbing clouds was simple during the 2016 monitoring.

Given the quasi-monotonic and severe decrease in the flux density of \PKS1830\ during our six-month monitoring (by  $\sim 80$\% ), the short (timescale of the order of one week) and regular variations of the absorption spectra, and the apparent coherence (periodicity?) of the variations over $\gtrsim 200$~days seen toward both lensed images, we conclude that they are most likely due to intrinsic changes in the background continuum distribution, that is, the activity of the quasar, rather than to micro-lensing events. Eventually, the different scenarios (helical jet or lensing-induced variations) could be tested with long-term (i.e., more than one year long) monitoring of the absorption variation and its potential periodicity (e.g., with ALMA) and continuum structural changes (e.g., with VLBI), although flaring or micro- and milli-lensing events could occasionally confuse a periodic signal induced by the jet helicity.

\subsubsection{Variations on short timescales}

We now consider variations on short timescales ---that is, between two consecutive observations at $t_1$ and $t_2$--- resulting from intrinsic morphological changes in the blazar. Accordingly, we assume that the characteristic size scale, a causal region of size $c(t_2-t_1)$, is small compared to the overall continuum emission. 

We can then introduce the continuum distribution $i_2(x,y,t_2) = i_1(x,y,t_1) + \delta i(x,y)$, where $\delta i(x,y) = 0$ outside of the small region of the continuum that was indeed affected by changes between $t_1$ and $t_2$.
We do not need to assume any specific shape for this region, except that it is strictly included within a circle of radius $c(t_2-t_1)$ or $\theta_E$ centered toward the direction $x_c, y_c$ in the sky. We express the new normalized absorption spectrum $S_2$ from Eq.\,\ref{eq:Inu}:
\begin{multline}
  S_2 = \frac{1}{I_1+\delta I} \iint (i_1(x,y)+\delta i(x,y)) {\rm e}^{-\tau (x,y)} dxdy.
\end{multline}
\noindent Rewriting the equation for $\delta I/I_1<<1$, and neglecting the second-order terms in $\delta I/I_1$, we get:
\begin{equation}
 \Delta S = S_2 -S_1 = \frac{1}{I_1} \iint \left ( \delta i - \frac{\delta I}{I_1} i_1 \right ) {\rm e}^{-\tau} dx dy.
\end{equation}
\noindent Furthermore, we assume that the absorbing material illuminated by the continuum region $\delta i(x,y)$ has homogeneous opacity $\tau_c = \tau(x_c,y_c)$. After simplifying the integral summation, we obtain:
\begin{equation} \label{eq:deltaS}
 \Delta S = \frac{\delta I}{I_1} \left ( {\rm e}^{-\tau_c} - \langle {\rm e}^{-\tau} \rangle _i \right ).
\end{equation}
\noindent As already mentioned, $\langle {\rm e}^{-\tau} \rangle _i$ is the continuum intensity-weighted average of the exponential of the opacity distribution. Therefore, in order to observe variations between normalized absorption spectra, the necessary condition is that the opacity of the absorbing region behind which the continuum changed is different from the overall average opacity; in other words, that the absorbing material has substructures at scales smaller than the continuum illumination solid angle ($\Omega_{abs}$).

Because the function ${\rm e}^{-\tau}$ takes values strictly between 0 and 1, the most favorable cases to detect variations according to Eq.\,\ref{eq:deltaS} are: when  $\langle {\rm e}^{-\tau} \rangle _i \sim 1$ and ${\rm e}^{-\tau_c} \sim 0$, that is when a small and optically thick region is suddenly illuminated over an absorbing medium of mostly very low opacity; or when $\langle {\rm e}^{-\tau} \rangle _i \sim 0$ and ${\rm e}^{-\tau_c} \sim 1$, that is when the sudden change of continuum occurs behind an optically thin region, while previously the bulk of the absorption concerned optically thick material. The maximum possible change in the difference of normalized absorption spectra is $\delta I/I_1$.

\subsection{Correlation H$_2$O -- CH}

\begin{figure*}[h] \begin{center}
\includegraphics[width=\textwidth]{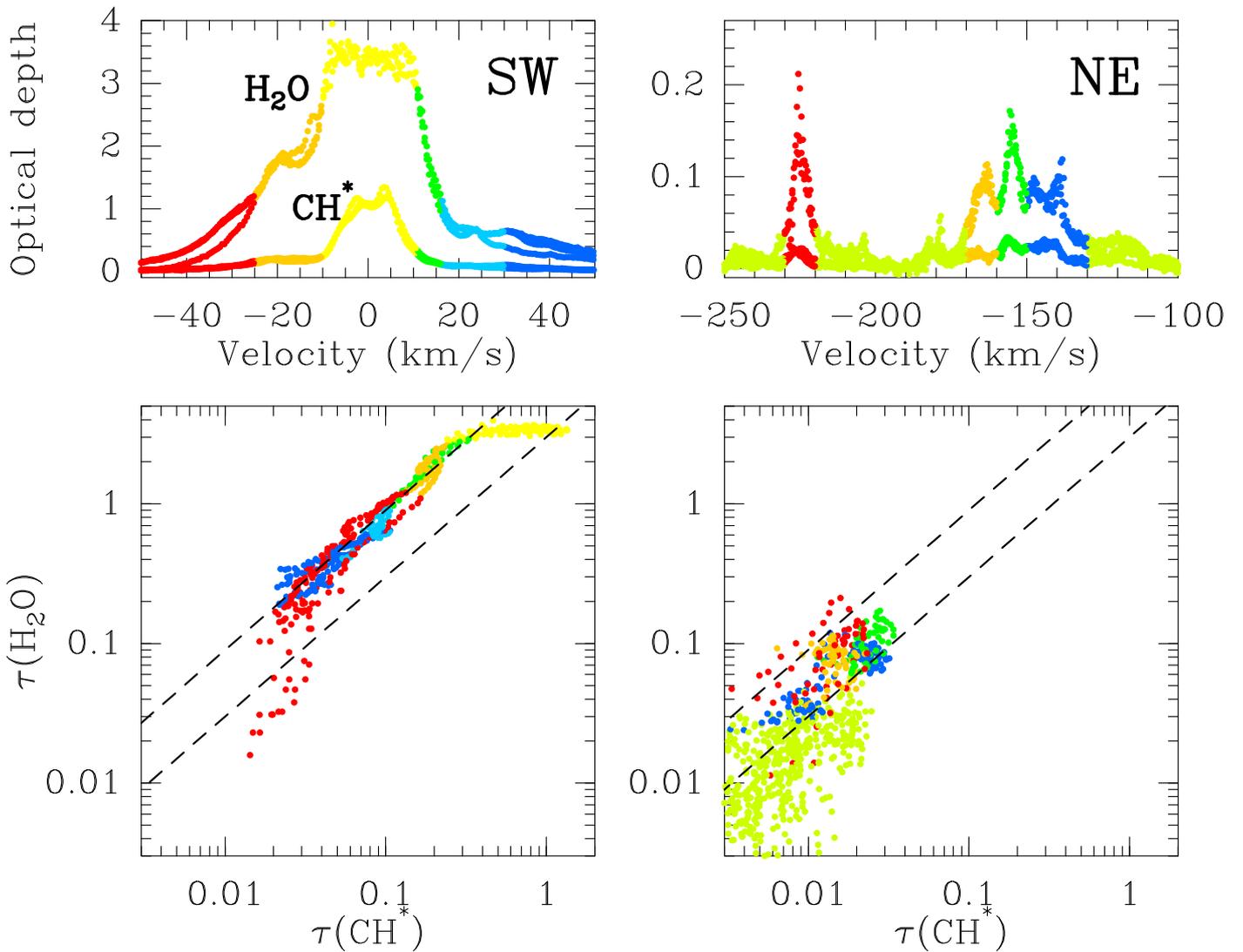}
\caption{Correlation between the optical depths of the ortho-H$_2$O $1_{1,0}-1_{0,1}$ line and of the deconvolved profile of CH~3/2-1/2 (given for an equivalent hyperfine component with normalized intensity $S_{ul}=1$) along the SW ({\em left}) and NE ({\em right}) lines of sight, respectively. In both lines of sight, we assume a source-covering factor of \fc=1. The colors encode the velocity intervals shown on the line profiles (top). The dashed lines correspond to opacity ratios $\tau$(H$_2$O)/$\tau$(CH$^*$)=3 and 9, respectively.}
\label{fig:corr-H2O-CH-fc1.0}
\end{center} \end{figure*}

Assuming that the absorbing material covers the background continuum source   entirely (\fc=1), we can directly convert the absorption profile into optical depth from Eq.\,\ref{eq:Inu2}. With this assumption, it is also possible to deconvolve the hfs pattern of the CH doublets, for example using a Clean algorithm method assuming that the relative intensities of the hfs components follow their statistical weights. Accordingly, we can investigate the optical-depth correlation between the H$_2$O and CH lines, as shown in Fig.\ref{fig:corr-H2O-CH-fc1.0} for both the SW and NE lines of sight.

In the Milky Way, both H$_2$O and CH are found to be excellent tracers of the H$_2$ along diffuse molecular sightlines. \cite{she08} derive a quasi-linear relationship between the column densities of CH and H$_2$, with an average abundance ratio of [CH]/[H$_2$]~$= 3.5 _{-1.4}^{+2.1} \times 10^{-8}$ along diffuse molecular sightlines and dark clouds. For water, \cite{fla13} also find a remarkably constant abundance ratio relative to H$_2$ in Galactic translucent clouds, with [H$_2$O]/[H$_2$]~$= (4.8 \pm 0.3) \times 10^{-8}$. Taking these relative abundances with their uncertainties, and assuming (i) that the relative abundances of H$_2$O and CH for the absorbing material in the \PKS1830\ absorber are similar to the Galactic clouds, (ii) an ortho-to-para ratio of three for H$_2$O, and (iii) that the molecules are at excitation equilibrium with cosmic-microwave-background photons ($\Tcmb = 5.14$~K at $z=0.89$, \citealt{mul13}), we would expect opacity ratios $\tau$(H$_2$O)/$\tau$(CH$^*$) of between 3 and 9 (at the 1$\sigma$ confidence level), CH$^*$ being the normalized CH spectrum (equivalent to a hyperfine structure component of relative strength $S_{ul}=1$) obtained after deconvolution of the hyperfine structure of the two CH $\Lambda$-doublets.


Discarding the center of the SW line profile (velocities between $\sim -10$ and +10~\kms), for which the water line is heavily saturated and has $\fc \sim 95$\% (which is therefore not consistent with our working assumption $\fc=1)$, we find that the H$_2$O/CH opacity ratios along the SW line of sight are roughly around 9 (see also fit results in Table~\ref{tab:FitOxygenRatios}), that is, in the upper part of Galactic ratios. Along the NE line of sight, the $\tau$(H$_2$O)/$\tau$(CH$^*$) opacity ratios are somewhat lower than in the SW line of sight, although the spread may partly be due to lower signal-to-noise ratios. These ratios are nevertheless still within the range of values observed in Milky Way translucent clouds. Based on the overall similarity between the $\tau$(H$_2$O)/$\tau$(CH$^*$) ratios in the \PKS1830\ absorber and in Galactic clouds, and except for the heavily saturated part of the SW water line, we do not find evidence for large opacity effects in other velocity components. Should the source-covering factor be lower than unity for some velocity components, the true optical depth of the water line would rise faster than that of CH, and the points in Fig.\,\ref{fig:corr-H2O-CH-fc1.0} would deviate from the straight correlation in the same way as for the saturated region near $v=0$~\kms. Therefore, regarding H$_2$O and CH, it seems that the chemical conditions in the different velocity components along the SW line of sight are consistent with a relatively constant H$_2$O/CH abundance ratio.

Moreover, the differences in H$_2$O and CH$^*$ opacity ratios between the SW and NE lines of sight follow a similar trend to that observed in the Milky Way between clouds in the central molecular zone (CMZ) and the Galactic disk, for which \cite{son13} found a water abundance three to five times higher in the CMZ. The cosmic-ray (CR) ionization rate deduced from a simple analysis of the OH$^+$ and H$_2$O$^+$ ions also follows a similar trend between the CMZ and disk of the Milky Way \citep{ind15} on one hand, and the SW and NE lines of sight toward \PKS1830 \citep{mul16b} on the other. As these ions are precursors of H$_2$O, the somewhat larger H$_2$O abundance could be related to a higher CR ionization rate.

\section{Summary and conclusions} \label{sec:summary}

We monitored the lensed quasar \PKS1830\ with ALMA over a time span of six months in order to investigate the variability of the submm continuum emission of the blazar and of its foreground molecular absorption. We summarize our results as follows. First, for the continuum activity:

\begin{itemize}
\item The submm light curve shows a dramatic and nearly monotonic decrease of 80\% in the flux density during our monitoring.
\item In contrast, the flux-density ratios between the two lensed images of the quasar show only mild variations of less than 10\%. There are signs of variations of the polarization properties, which manifest as fluctuations in the differential polarization between the two images. Toward the end of the monitoring, the quasar seems to have reached a stable stage in its polarization variability.
\item The nearly featureless light curve prevents us from obtaining a strong constraint on the time delay between the two images. Looking backwards at the previous measurements of the flux-density ratios, it appears that it is also difficult to precisely determine the differential magnification ratio of the system, as the long-term evolution of the flux-density ratios suggests milli-lensing events on timescales of several years.
\item We propose a relatively simple parametric model of a single ballistic synchrotron-cooling plasmon to account for the evolution of the continuum emission during the monitoring.
\item From this parametric model, we also obtain a new measurement of the time delay between the NE and SW images of $\Delta t = 25 \pm 3$~days, and of their differential magnification, namely $1.24 \pm 0.04$, both of which are consistent with values provided by modeling of the system and its lensing geometry.
\end{itemize}

For the absorption lines arising in the $z=0.89$ lens--absorber:

\begin{itemize}
\item The absorption line profiles of H$_2$O (saturated with optical depth $>1$) and CH (peak opacity $\tau \sim 1$) reveal clear time variations (on average up to $\sim1$\% of the continuum level per day for the water line) between spectra taken 1--3 weeks apart.
\item A principal component analysis of the absorption spectra toward both images of \PKS1830\ reveals a seemingly wavy behavior of the variations during the monitoring, with an apparent period of $\sim 230$~days, although this timescale is longer than the time span of our survey and would therefore need to be confirmed with a longer monitoring. It is tempting to relate this period to the jet precession period of approximately one year in the scenario of a helical jet proposed by \cite{nai05}. This would provide a natural explanation for the time variability of the absorption line.
\item Combining all the spectra along the SW image, we obtain a detection of the rare isotopolog $^{13}$CH, with a $^{12}$CH/$^{13}$CH ratio of $\sim 150$. From complementary data, we also measure oxygen isotopic ratios and derive $^{16}$O/$^{18}$O $= 65.3 \pm 0.7$ and $^{18}$O/$^{17}$O $= 11.5 \pm 0.5$.
\item We discover a remarkable, broad but shallow absorption trough of 500~\kms\ in width observed in the water line toward the SW image, which could be the signature of extra-planar molecular gas.
\item We explore the chemical correlation between CH and H$_2$O using their optical depth ratio and find a H$_2$O/CH ratio comparable to that observed in Milky Way clouds.
\end{itemize}
  
The lens--absorber system toward \PKS1830\ acts as a powerful cosmological microscope, allowing us to investigate the coupling between the quasar activity and the variations of the absorption pattern. In particular, whether the absorption variability is periodic ---and is therefore most likely connected with the precession of the quasar's helical jet--- or irregular ---and is connected to occasional flaring or micro-/milli-lensing events--- remains to be confirmed.

\begin{acknowledgement}
This paper makes use of the following ALMA data:
ADS/JAO.ALMA\#2011.0.00405.S (observations of H$_2$O and CH in April 2012),
ADS/JAO.ALMA\#2012.1.00056.S (observations of H$_2$O and CH in May 2014),
ADS/JAO.ALMA\#2013.1.00020.S (observations of H$_2$O, CH, H$_2^{18}$O, and H$_2^{17}$O in July 2014),
ADS/JAO.ALMA\#2015.1.00075.S (2016 monitoring of H$_2$O and CH),
ADS/JAO.ALMA\#2018.1.00051.S (observations of H$_2^{18}$O in 2019),
ADS/JAO.ALMA\#2021.A.00028.S (observations of H$_2^{18}$O and H$_2^{17}$O in 2022).
ALMA is a partnership of ESO (representing its member states),
NSF (USA) and NINS (Japan), together with NRC (Canada), NSC and ASIAA (Taiwan), and KASI
(Republic of Korea), in cooperation with the Republic of Chile.
The Joint ALMA Observatory is operated by ESO, AUI/NRAO and NAOJ.
This work has been partially supported by Generalitat Valenciana (GenT Project CIDEGENT/2018/021 and grant ASFAE/2022/018) and the MICINN (Research Project PID2019-108995GB-C22 and the Astrophysics and High Energy Physics programme, with funding from European Union NextGenerationEU, PRTR-C17I1).
This research has made use of NASA's Astrophysics Data System.
\end{acknowledgement}


\begin{appendix}

\section{Spectroscopic data}

Spectroscopic data for the transitions of H$_2$O and CH isotopologs observed in this study are given in Tab.\ref{tab:spectro}.

\begin{table*}[ht!]
\caption{Spectroscopic parameters for transitions of H$_2$O and CH isotopologs observed in this study.}
\label{tab:spectro}
\begin{center} \begin{tabular}{cccccc}
\hline
Species & Transition & Rest Freq.  &   Redshifted Freq. & $S_{ul}$ &  $E_{low}$ \\ 
        &   & (GHz)       &  (GHz)            &         &  (K) \\
\hline
ortho-H$_2$O  & $J_{K_{\rm a},K_{\rm c}} = 1_{1,0} - 1_{0,1}$ & 556.9359877 & 295.328 & 4.5 & 0.0 \\
ortho-H$_2^{18}$O  & $J_{K_{\rm a},K_{\rm c}} = 1_{1,0} - 1_{0,1}$ & 547.676440 & 290.418 & 4.5 & 0.0 \\
ortho-H$_2^{17}$O  & $J_{K_{\rm a},K_{\rm c}} = 1_{1,0} - 1_{0,1}$ & 552.020960 & 292.722 & 4.5 & 0.0 \\
\hline
CH & $(J^p,F)= (3/2^+,1)-(1/2^-,1)$ & 532.7215886 & 282.488 & 0.17 & 0.2 \\
   & $(J^p,F)= (3/2^+,2)-(1/2^-,1)$ & 532.7238893 & 282.489 & 0.83 & 0.2 \\
   & $(J^p,F)= (3/2^+,1)-(1/2^-,0)$ & 532.7932746 & 282.526 & 0.33 & 0.2 \\
   & $(J^p,F)= (3/2^-,2)-(1/2^+,1)$ & 536.7610463 & 284.630 & 0.83 & 0.0 \\
   & $(J^p,F)= (3/2^-,1)-(1/2^+,1)$ & 536.7818563 & 284.641 & 0.17 & 0.0 \\
   & $(J^p,F)= (3/2^-,1)-(1/2^+,0)$ & 536.7955695 & 284.648 & 0.33 & 0.0 \\
\hline    
$^{13}$CH & $(J^p,F_1,F)= (3/2^+,1,1/2)-(1/2^-,1,3/2)$ &  531.859975 &    282.031 &     0.04 &  0.2 \\ 
         & $(J^p,F_1,F)= (3/2^+,1,3/2)-(1/2^-,1,3/2)$ &  531.862711 &    282.033 &     0.19 &  0.2 \\ 
         & $(J^p,F_1,F)= (3/2^+,1,1/2)-(1/2^-,1,1/2)$ &  531.910901 &    282.058 &     0.05 &  0.2 \\ 
         & $(J^p,F_1,F)= (3/2^+,1,3/2)-(1/2^-,1,1/2)$ &  531.913471 &    282.060 &     0.07 &  0.2 \\ 
         & $(J^p,F_1,F)= (3/2^+,2,3/2)-(1/2^-,1,3/2)$ &  532.083360 &    282.150 &     0.11 &  0.2 \\ 
         & $(J^p,F_1,F)= (3/2^+,2,5/2)-(1/2^-,1,3/2)$ &  532.086251 &    282.151 &     1.00 &  0.2 \\ 
         & $(J^p,F_1,F)= (3/2^+,2,3/2)-(1/2^-,1,1/2)$ &  532.134740 &    282.177 &     0.55 &  0.2 \\ 
         & $(J^p,F_1,F)= (3/2^+,1,1/2)-(1/2^-,0,1/2)$ &  532.224939 &    282.225 &     0.25 &  0.1 \\ 
         & $(J^p,F_1,F)= (3/2^+,1,3/2)-(1/2^-,0,1/2)$ &  532.227528 &    282.226 &     0.41 &  0.1 \\ 
         & $(J^p,F_1,F)= (3/2^+,2,3/2)-(1/2^-,0,1/2)$ &  532.448670 &    282.343 &     0.01 &  0.1 \\ 

         & $(J^p,F_1,F)= (3/2^-,1,3/2)-(1/2^+,0,1/2)$ &  536.005094 &    284.229 &     0.49 &  0.0 \\ 
         & $(J^p,F_1,F)= (3/2^-,1,1/2)-(1/2^+,0,1/2)$ &  536.024969 &    284.240 &     0.17 &  0.0 \\ 
         & $(J^p,F_1,F)= (3/2^-,1,3/2)-(1/2^+,1,3/2)$ &  536.026643 &    284.241 &     0.17 &  0.0 \\ 
         & $(J^p,F_1,F)= (3/2^-,1,3/2)-(1/2^+,1,1/2)$ &  536.037944 &    284.247 &     0.01 &  0.0 \\ 
         & $(J^p,F_1,F)= (3/2^-,1,1/2)-(1/2^+,1,3/2)$ &  536.046477 &    284.251 &     0.04 &  0.0 \\ 
         & $(J^p,F_1,F)= (3/2^-,1,1/2)-(1/2^+,1,1/2)$ &  536.057826 &    284.257 &     0.13 &  0.0 \\ 
         & $(J^p,F_1,F)= (3/2^-,2,3/2)-(1/2^+,0,1/2)$ &  536.099484 &    284.279 &     0.01 &  0.0 \\ 
         & $(J^p,F_1,F)= (3/2^-,2,5/2)-(1/2^+,1,3/2)$ &  536.101144 &    284.280 &     1.00 &  0.0 \\ 
         & $(J^p,F_1,F)= (3/2^-,2,3/2)-(1/2^+,1,3/2)$ &  536.121035 &    284.291 &     0.13 &  0.0 \\ 
         & $(J^p,F_1,F)= (3/2^-,2,3/2)-(1/2^+,1,1/2)$ &  536.132344 &    284.297 &     0.53 &  0.0 \\ 
\hline
\end{tabular}
\tablefoot{Sky frequencies are calculated for $z$=0.88582. 
For H$_2$O, the frequency is taken from \cite{caz09}.
For CH, frequencies are taken from \cite{tru14} and line relative intensities ($S_{ul}$) from \cite{nij12}. Each rotational $J$-level is split into two opposite parity states ($p=+,-$) by $\Lambda$-doubling. The hydrogen nuclear spin ($I=1/2$) further splits each level into hyperfine components ${\bf F} = {\bf J}+{\bf I}$.
For $^{13}$CH, spectroscopic data are taken from \cite{hal08} and the JPL molecular spectroscopy database. Both nuclei have spin angular momentum. $J$ couples with the $^{13}$C spin ($I_1=1/2$) ${\bf F_1} = {\bf J}+{\bf I_1}$. Then, $F_1$ couples with the hydrogen spin ($I_2=1/2$): ${\bf F} = {\bf F_1}+{\bf I_2}$.}
\end{center} \end{table*}

\section{Additional material}

\begin{figure}[h] \begin{center}
\includegraphics[width=8.8cm]{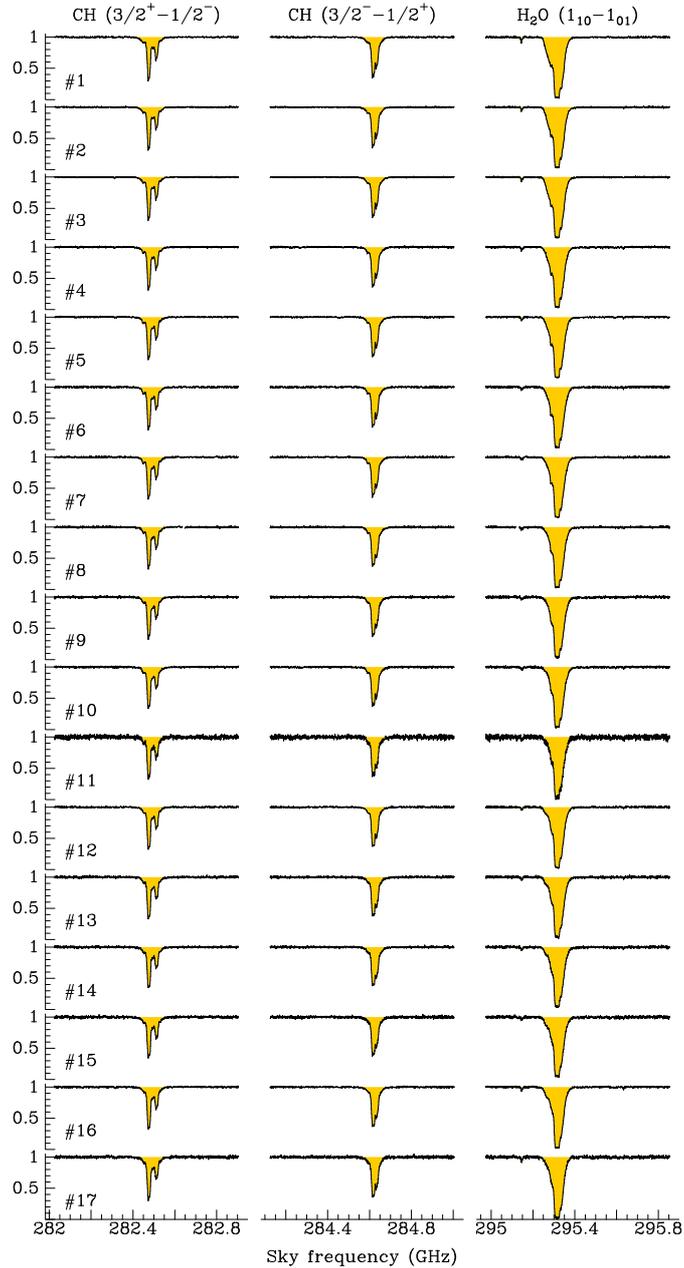}
\caption{Overview of the continuum-normalized absorption spectra of the CH $J^p = 3/2^+-1/2^-$ ({\em left}), CH $J^p = 3/2^--1/2^+$ ({\em middle}), and ortho-H$_2$O $J_{K_a,K_c} = 1_{1,0}-1_{0,1}$ ({\em right}) lines toward the SW image of \PKS1830, for each visit of the 2016 monitoring. The visit numbers \#i correspond to the entries in Table~\ref{tab:contdata-c3-ave}.}
\label{fig:monitoring-specSW}
\end{center} \end{figure}

\begin{figure}[h] \begin{center}
\includegraphics[width=8.8cm]{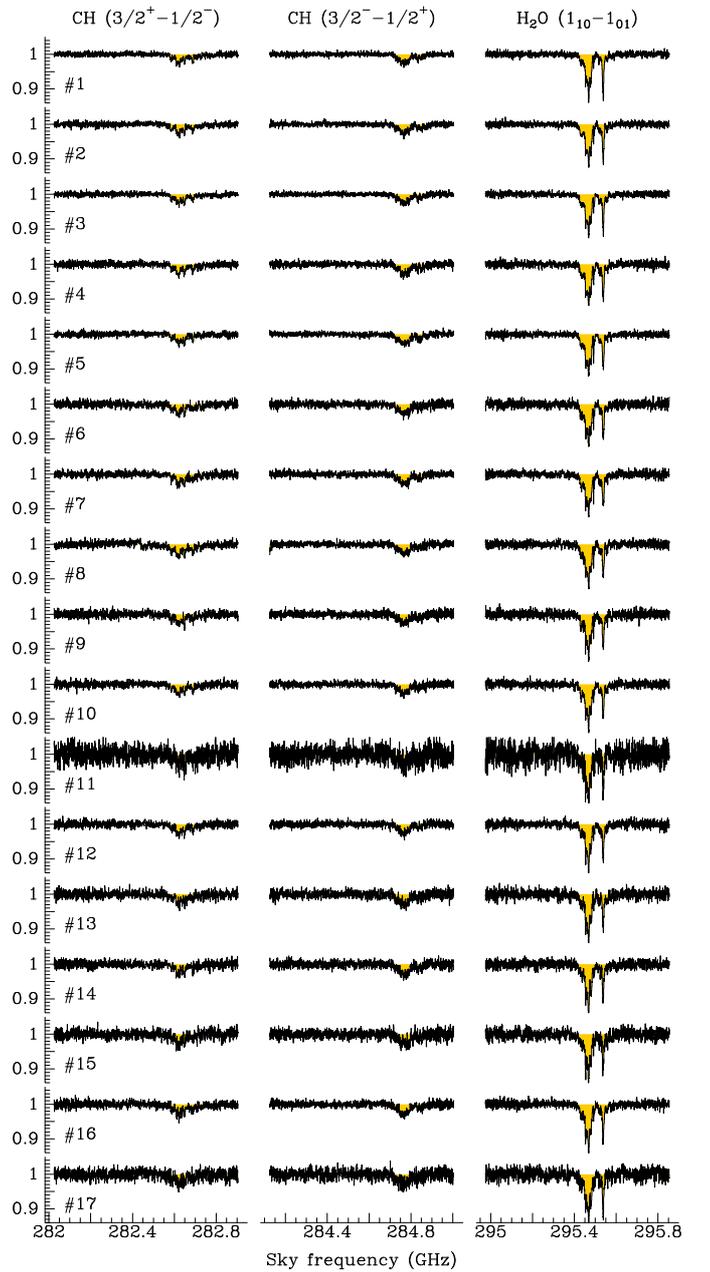}
\caption{Same as Fig.\,\ref{fig:monitoring-specSW}, but toward the NE image of \PKS1830.}
\label{fig:monitoring-specNE}
\end{center} \end{figure}

We provide an overview of the ALMA continuum-normalized absorption spectra of H$_2$O and CH toward the SW and NE image of \PKS1830 during our monitoring in Figs.\,\ref{fig:monitoring-specSW} and \ref{fig:monitoring-specNE}, respectively.

\end{appendix}
\end{document}